\documentclass{JHEP3}
\usepackage[english]{babel}
\usepackage{cite}

\usepackage{epsf}
\usepackage{amssymb}
\usepackage{amsmath}
\usepackage{amsfonts}
\usepackage{psfrag,epsfig,graphicx,graphics}

\newcommand{\g}{\gamma}

\newcommand{\kb}{\underline{k}}

\def\slashchar#1{\setbox0=\hbox{$#1$}
   \dimen0=\wd0
   \setbox1=\hbox{/} \dimen1=\wd1
   \ifdim\dimen0>\dimen1
      \rlap{\hbox to \dimen0{\hfil/\hfil}}
      #1
   \else
      \rlap{\hbox to \dimen1{\hfil$#1$\hfil}}
      /
   \fi}

\def\ps{\slashchar{p}}

\def\es{\slashchar{\epsilon}}
\newcommand{\ks}{\mbox{$\slashchar{k}$}}

\def\tr{{\rm tr}}

\def\no{\noindent}

\def\bei{\begin{itemize}}
\def\ei{\end{itemize}}

\def\beeq{\begin{eqnarray}} 
\def\beqa{\begin{eqnarray}}
\def\bea{\begin{eqnarray}}

\def\eea{\end{eqnarray}}
\def\eqa{\end{eqnarray}}
\def\eeeq{\end{eqnarray}}

\def\eqar{\end{array}}
\def\beqar{\begin{array}}

\def\beas{\begin{eqnarray*}}
\def\beqas{\begin{eqnarray*}}

\def\eqas{\end{eqnarray*}}
\def\eeas{\end{eqnarray*}}

\def\beq{\begin{equation}} 
\def\be{\begin{equation}}

\def\ee{\end{equation}}
\def\eq{\end{equation}}
\def\eeq{\end{equation}}

\def\beqd{\begin{displaymath}}
\def\eeqd{\end{displaymath}}
\def\eqd{\end{displaymath}}

\def\beeq{\begin{eqnarray}} \def\eeeq{\end{eqnarray}}

\newcommand{\fin}{\end{document}}

\title{Resumming  soft and collinear contributions in deeply virtual Compton scattering}
\author{T. Altinoluk\\
CPhT, \'Ecole Polytechnique,
CNRS, F-91128 Palaiseau,     France \\
Email: \email{tolga.altinoluk@cpht.polytechnique.fr}}
\author{ B.~Pire \\
CPhT, \'Ecole Polytechnique,
CNRS, F-91128 Palaiseau,     France \\
Email: \email{Bernard.Pire@cpht.polytechnique.fr}}
\author{ L. Szymanowski\\
National Center for Nuclear Research (NCBJ), Warsaw, Poland\\
Email: \email{Lech.Szymanowski@fuw.edu.pl}}
\author{S. Wallon\\
LPT, Universit{\'e} Paris-Sud, CNRS, 91405, Orsay, France {\em \&} \\
UPMC Univ. Paris 06, facult\'e de physique, 4 place Jussieu, 75252 Paris Cedex 05, France\\
Email: \email{wallon@th.u-psud.fr}}

\abstract{
We calculate  the quark coefficient function $T^q(x,\xi)$  that enters the
factorized amplitude for deeply virtual Compton scattering (DVCS) at all order in a
soft and collinear gluon approximation, focusing on the leading double logarithmic
behavior in $(x\pm\xi)$, where $x\pm\xi$ is the
light cone momentum fraction of the incoming/outgoing quarks.
We show that the dominant part of the known one loop
result can be understood in an axial gauge as the result of a semi-eikonal
approximation to the box diagram. We then derive an all order result for the leading
contribution of the ladder diagrams and deduce a resummation formula valid in the
vicinity of the boundaries of the regions defining the energy flows of the
incoming/outcoming quarks, i.e. $x=\pm\xi$.  The resummed series results in a simple
closed expression.
}

\date{\today}
\begin{document}

\pagestyle{empty}
\newpage

\mbox{}

\pagestyle{plain}

\setcounter{page}{1}
\section{Introduction}
\label{Sec:Intro}

Since a decade, there has been much progress in the understanding of the three-dimensional content of the hadron, both from the theory and the experimental sides. Experimentally, this relies on several new electron facilities combining high luminosity and advanced detectors which allow for measuring with an impressive precision exclusive processes, including deep virtual Compton scattering (DVCS) and meson production. This lead to the first studies of non-perturbative non forward parton distributions, now called generalized parton distributions (GPDs), first in the fixed target experiment HERMES \cite{Airapetian:2001yk, Ellinghaus:2002bq, Airapetian:2006zr}, and then at H1 and ZEUS, using the dominance of the DVCS contribution at small $x_{Bj}$
\cite{Adloff:2001cn, Aktas:2005ty, Chekanov:2003ya}. Almost simultaneously, the DVCS contribution was measured at JLAB,  at CLAS \cite{Stepanyan:2001sm, Girod:2007jq} and  at Hall A \cite{MunozCamacho:2006hx}. From the theory side, the interest for hard exclusive processes started with the Leipzig group \cite{Mueller:1998fv}.  Several studies\footnote{For reviews, see Refs.~\cite{Guichon:1998xv, Goeke:2001tz, Diehl:2003ny, Belitsky:2005qn}} then set the basis of a consistent framework, called collinear factorization, to separate the 
short distance dominated partonic subprocesses and long distance hadronic matrix elements, at leading  and next-to-leading order for DVCS \cite{Radyushkin:1996nd, Radyushkin:1997ki, Belitsky:1997rh, Belitsky:1999sg, Ji:1998xh, Collins:1998be} and for hard electroproduction of  mesons \cite{Collins:1996fb, Radyushkin:1996ru, Belitsky:2001nq, Ivanov:2004zv}  and their timelike crossed versions, namely exclusive lepton pair production in photon or meson collisions with protons \cite{Berger:2001zn, Berger:2001xd, Pire:2008ea}. The future JLab-12 GeV and COMPASS-II program will provide soon bunches of data, giving a hope to get access to GPDs with a high degree of precision. There are indeed now intense activities \cite{Guidal:2008ie, Moutarde:2009fg, Guidal:2009aa, Moutarde:2012ht} to move from discovery era to precision physics. 

In order to extract the GPDs, a precise theoretical framework should be available, which should go beyond a pure leading logarithmic treatment both for evolution equations and for coefficient functions, with the expected increase of precision of future data. The aim of this paper is to study in detail the emergence of the leading contributions near the points $x =\pm \xi$ and to derive a resummed
formula for the coefficient function of DVCS, which could have a major phenomenological impact in future precise studies. A brief report on this result has been presented elsewhere \cite{Altinoluk:2012fb}.

\begin{figure}[h]
\centerline{\scalebox{1}{
\begin{tabular}{c}
\hspace{1.35cm}\raisebox{-.44 \totalheight}{
\psfrag{p}[cc][cc]{$\Gamma_2$}
\psfrag{q1}[cc][cc]{$q$}
\psfrag{gas}[cc][cc]{\raisebox{1cm}{$\hspace{-.3cm}\gamma^*$}}
\psfrag{g}[cc][cc]{\raisebox{1cm}{$\gamma$}}
\psfrag{H}[cc][cc]{$T^q$}
\psfrag{k1}[cc][Bc]{\raisebox{0cm}{$\hspace{-.8cm} (x+\xi) p_2$}}
\psfrag{k2}[cc][Bc]{\raisebox{0cm}{$\hspace{.6cm} (x-\xi) p_2$}}
\psfrag{S}[cc][cc]{${\cal S}$}
\psfrag{U}[cc][cc]{${\cal U}$}
\includegraphics[height=5cm]{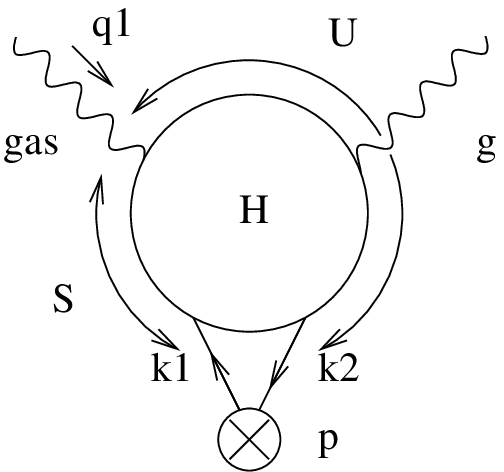}} 
\\ 
%
\psfrag{pl}[rc][Br]{}
\psfrag{pr}[cc][Bc]{}
\psfrag{pp}[rc][rc]{}
\psfrag{ppp}[cc][lc]{}
\psfrag{pmq}[cc][cc]{$+ \, -$}
\psfrag{mu}[cc][cc]{$+$}
\psfrag{n}[cc][cc]{$\Gamma_1$}
\psfrag{S}[cc][cc]{$F^q$}
\hspace{1.8cm}
\raisebox{-1 \totalheight}{\includegraphics[height=4cm]{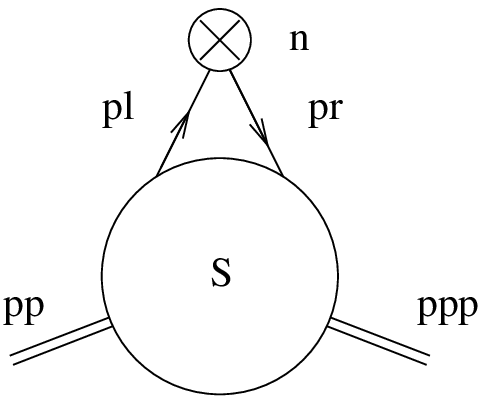}} 
\end{tabular}
}}
\caption{Factorization of the DVCS amplitude in the hard regime. The crossed-blob denote a set of $\Gamma$ matrices. In this paper  $\Gamma_i = \ps_i$\,.  In the above (hard) part, called coefficient function, the lines entering and exiting the crossed blob carry spinor and color indices but do not propagate any momentum. The corresponding momenta are on-shell.  
}
\label{Fig:factGPD}
\end{figure}

The  amplitude for the DVCS process
\begin{equation}
\gamma^{(*)}(q) N(p) \to \gamma(q') N'(p')\,,
\label{reaction}
\end{equation}
with a large  virtuality $q^2 =-Q^2$, factorizes at the leading twist 2 level in terms of perturbatively calculable coefficient functions $C(x,\xi, \alpha_s) $ and  GPDs $F(x,\xi,t)$, where the scaling variable in the generalized Bjorken limit is the skewness $\xi$ defined as 
\begin{equation}
\xi = \frac{Q^2}{(p+p')\cdot(q+q')}\,.
\label{eq:skewnessdef}
\end{equation}
%
Hereafter, we only consider quark exchange. After proper renormalization, this quark contribution to the symmetric part
of the factorized Compton scattering amplitude illustrated in Fig.~\ref{Fig:factGPD} reads 
\begin{eqnarray}
\mathcal{A}^{\mu\nu} = g_T^{\mu\nu}\int_{-1}^1 dx 
\left[
\sum_q^{n_F} T^q(x) F^q(x)
\right]\,,
\label{eq:factorizedamplitude}
\end{eqnarray}
where the  quark coefficient function $T^q$ read \cite{Pire:2011st} :
\begin{eqnarray}
&&\hspace{-.9cm}T^q=C_{0}^q +C_1^q +C_{coll}^q \log \frac{|Q^2|}{\mu^2_F}  \,,\label{tq}\\
&&\hspace{-.9cm}C_0^q =e_q^2\left(\frac{1}{x-\xi+i\varepsilon} \,-\, (x \to -x)  \right) \,, \label{C0} \\
&&\hspace{-.9cm}C_1^q \!=\!\frac{e_q^2\alpha_SC_F}{4\pi(x-\xi+i\varepsilon)}\!
\bigg[\!\!\log^2\!\!\bigg(\frac{\xi-x}{2\xi}-i\varepsilon \bigg)-9-\frac{3(\xi-x)}{\xi+x}\log\bigg(\frac{\xi-x}{2\xi}-i\varepsilon\!\!\bigg)\!\bigg]\!-\!(x \to -x) \,, \\
 \label{C1}
&&\hspace{-.9cm}C^q_{coll}=\frac{e_q^2\alpha_SC_F}{4\pi(x-\xi+i\varepsilon)}\bigg[3+2\log\bigg(\frac{\xi-x}{2\xi}-i\varepsilon\bigg)\bigg] -(x\rightarrow-x) \,.\label{Ccoll}
\end{eqnarray}
The first (resp. second) terms in Eqs.~(\ref{C0}) and (\ref{C1}) correspond to the $s-$channel (resp. $u-$channel) class of diagrams. One goes from the $s-$channel to the  $u-$channel by the interchange of the photon attachments. Since these two contributions are obtained from one another by a simple ($x\leftrightarrow -x$) interchange, we will restrict in the following mostly to the discussion of the former class of diagrams.

Eqs.~(\ref{C0}) and (\ref{C1}) show that among the corrections of $O(\alpha_s)$ to the coefficient function  the terms of order $[\log^2(\xi \pm x)]/(x\pm\xi)$ play an important role in the region of small $(\xi \pm x)$,  i.e. in the vicinity of the boundary between the so-called ERBL and DGLAP domains where the evolution equations of GPDs take distinct forms. We here scrutinize these regions and demonstrate that they are dominated by soft-collinear singularities.

The source of these singularities can be understood in the following way. 
In our analysis we expand any momentum in the Sudakov basis $p_1$, $p_2$, where $p_2$ is the light-cone direction of the two incoming and outgoing partons ($p_1^2=p_2^2=0$, $2 p_1 \cdot p_2 =s = Q^2/2\xi $), as 
\beq
k = \alpha \, p_1 + \beta \, p_2 + k_\perp\,.
\label{def-Sudakov}
\eq
 In this basis,  
\beq
q=p_1-2 \, \xi \, p_2 \quad {\rm and} \quad p_1 \equiv q'\,.
\label{def-q}
\eq
Now,
the Mandelstam variables ${\cal S}$ and ${\cal U}$ for the coefficient function
illustrated in the upper part of Fig.~\ref{Fig:factGPD} read
\beq
{\cal S}= \frac{x-\xi}{2 \xi} \, Q^2  \quad {\rm and}   \qquad {\cal U}= - \frac{x+\xi}{2 \xi} \, Q^2 \,.
\label{def-S-U}
\eq
Although the usual collinear approach is based on a single-scale analysis, where the only large scale is provided by $Q^2$, in the special kinematical limit where $x \to \xi$ (resp.  $x \to -\xi$), the Mandelstam variable ${\cal S}$ (resp.  ${\cal U}$) becomes parametrically small with respect to $Q^2$. We thus turn to a two scale problem, in a similar way as for the $x \to x_{Bj}$ limit of deep inelastic scattering (DIS) on a parton of longitudinal momentum fraction $x$.
In this limit, large terms of type $[\alpha_s \log^2 (\xi \pm x)]/(x \pm \xi)$ should appear, calling for a resummation of these threshold singularities,
similarly to the resummation of large $x_{Bj}/x$ coefficient functions in DIS \cite{Sterman:1986aj, Catani:1989ne}. As for DIS, the resummation which we now perform  is due to the combination of soft and collinear singularities. The main complication with respect to DIS is due to the non forward kinematics of DVCS. Our treatment relies on a diagramatic analysis, which we now explain.   

\begin{figure}
\centerline{\includegraphics[width=8cm]{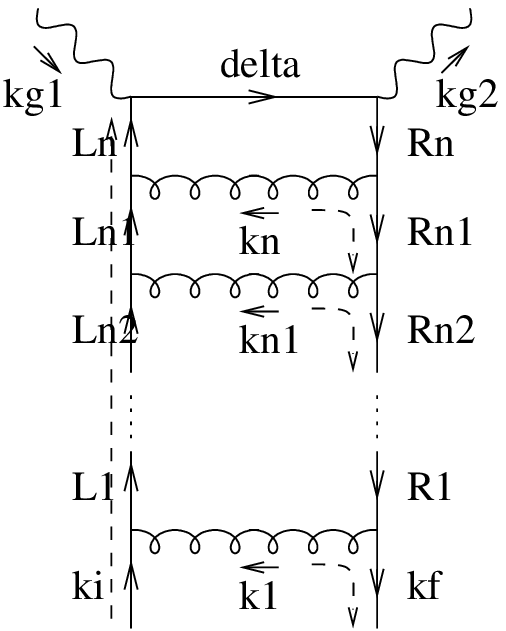}  
}
\caption{The ladder diagrams which contribute in the light-like gauge to the leading  $\alpha_s^n \ln^{2 n}(\xi-x)/(x-\xi)$ terms in the perturbative expansion of the    DVCS  amplitude. The $p_2$ and $\perp$ momentum components are indicated. The dashed lines show the dominant momentum flows along the $p_2$ direction.} 
\label{Fig:n-loop}
\end{figure}

We start our analysis by observing that in the same spirit as for evolution equations, the extraction of the soft-collinear singularities which dominate the amplitude in the limit $x \to \pm \xi$ is made easier when using the light-like gauge $p_1 \cdot A=0$. We argue that in this gauge the amplitude is dominated by ladder-like diagrams, illustrated in Fig.~\ref{Fig:n-loop}. 

We now restrict our study to the case   $x \to +\xi$. The dominant kinematics is given by a strong ordering both in longitudinal and transverse momenta, according to
\begin{eqnarray}
&& \hspace{-.3cm}x \sim \xi\gg \vert \beta_1\vert  \sim \vert x-\xi \vert \gg \vert x-\xi +\beta_1\vert \sim \vert \beta_2 \vert \gg \cdots \nonumber \\
&&\hspace{-.4cm}
\cdots \gg   \vert x-\xi +\beta_1 +\beta_2 -\cdots+ \beta_{n-1} \vert  \sim  \vert \beta_n  \vert  , 
\label{kinematics_beta}
\\
&& \vert k_{\perp 1}^2\vert  \ll \vert k_{\perp 2}^2\vert  \ll \cdots \ll \vert k_{\perp n}^2\vert  \ll s \sim Q^2 \,,
\label{kinematics_k}
\\
&&  \vert \alpha_1 \vert  \ll \cdots \ll  \vert \alpha_n \vert  \ll 1\,,
\label{kinematics_alpha}
\end{eqnarray}
where $ \alpha_i$ and $\beta_i$ are momentum fractions  along the two dominant light cone directions of the exchanged
gluons and $k_{\perp i}$ 
their transverse momenta.
This ordering is related to the fact that the dominant double logarithmic contribution for each loop arises from the region of phase space where both soft and collinear singularities manifest themselves. In the limit $x \to \xi\,,$ the left fermionic line is a hard line, from which the gluons are emitted in an eikonal way, with a collinear ordering. For the right fermionic line, an eikonal approximation is not valid, since the dominant momentum flow along $p_2$ is from the gluon to the fermion. Even though this is the case, a collinear approximation can still be applied. This non-symmetric treatment of the whole diagram will be referred to as the semi-eikonal approximation.

\def\sca{.9}
\psfrag{delta}{\raisebox{.2cm}{\hspace{-1.3cm}\scalebox{\sca}{\begin{tabular}{c}$x-\xi+\beta_1 + \cdots +\beta_n\,,$\\
$k_{\perp 1}+ \cdots +k_{\perp n}$
\end{tabular}}}}

\psfrag{kg1}{\raisebox{.2cm}{\hspace{-.9cm}\scalebox{\sca}{$p_1 - 2 \xi \, p_2$}}}

\psfrag{kg2}{\raisebox{.2cm}{\hspace{.2cm}\scalebox{\sca}{$p_1$}}}


\psfrag{kn}{\hspace{-.7cm}\scalebox{\sca}{$\ \ \ \beta_{n}, \,k_{\perp\, n}$}}

\psfrag{kn1}{\hspace{-1cm}\scalebox{\sca}{$\ \ \ \beta_{n-1}, \,k_{\perp\, n-1}$}}

\psfrag{k1}{\hspace{-.6cm}\scalebox{\sca}{$\ \ \ \beta_1, \,k_{\perp 1}$}}

\psfrag{Ln}[R]{\hspace{-2.8cm} \scalebox{\sca}{\begin{tabular}{c}$x+\xi+\beta_1 + \cdots +\beta_n,$\\
$k_{\perp 1}+ \cdots +k_{\perp n}$\end{tabular}}}

\psfrag{Ln1}[R]{\hspace{-2.9cm} \scalebox{\sca}{\begin{tabular}{c}$x+\xi+\beta_1 + \cdots +\beta_{n-1},$\\$k_{\perp 1}+ \cdots +k_{\perp n-1}$\end{tabular}}}

\psfrag{Ln2}[R]{}

\psfrag{L1}{\hspace{-2.5cm}\scalebox{\sca}{$x+\xi+\beta_1, \,k_{\perp1}$}}

\psfrag{ki}[L]{\hspace{-.9cm}\scalebox{\sca}{$(x+\xi)\,p_2$}}

\psfrag{Rn}[R]{\hspace{2.9cm} \scalebox{\sca}{\begin{tabular}{c}$x-\xi+\beta_1 + \cdots +\beta_n,$\\
$k_{\perp 1}+ \cdots +k_{\perp n}$\end{tabular}}}

\psfrag{Rn1}[R]{\hspace{3.2cm} \scalebox{\sca}{\begin{tabular}{c}$x-\xi+\beta_1 + \cdots +\beta_{n-1},$\\
$ k_{\perp 1}+ \cdots +k_{\perp n-1}$\end{tabular}}}

\psfrag{Rn2}[R]{}

\psfrag{R1}{\hspace{-.2cm}\scalebox{\sca}{$x-\xi+\beta_1, \,k_{\perp 1}$}}

\psfrag{kf}[R]{\hspace{1cm} \scalebox{\sca}{$(x-\xi)\,p_2$}}



%

The issue related to the $i \epsilon$ prescription in Eq.~(\ref{C1}) is solved by computing the coefficient function in the unphysical region $\xi
> 1$. After analytical continuation to the physical region $0 \leq \xi \leq 1$, the physical prescription is then obtained through the shift $\xi \to \xi-i \epsilon$\,.

We denote by $K_n$ the contribution of a $n$-loop ladder to the coefficient  function, and define\footnote{Note that this normalization is fixed to match  with the definition of the coefficient function used in Eq.~(\ref{tq}), while  the $S$-matrix element is $i K_n$ at order $n$.}
\begin{eqnarray}
\label{def-Kn}
K_n = -\frac{1}4 e_q^2 \left(-i \, C_F \, \alpha_s \frac{1}{(2 \pi)^2} \right)^n I_n\,.
\end{eqnarray}
As mentioned earlier, Eq.~(\ref{def-Kn}) should be completed by inclusion of the $u$-channel class of diagrams. 


The paper is organized as follows.
In Sec.~\ref{Sec:1-order}, we perform a detailed analysis of the one-loop diagrams and extract the dominant contribution in  $[\alpha_s \log^{2}(\xi-x)]/(x-\xi)$. In Sec.~\ref{Sec:1-order-ladder}, 
we analyze the same one-loop
contribution but in the spirit of the semi-eikonal approximation
described above
 and show that the dominant contribution is indeed identical. In Sec.~\ref{Sec:2-order}, we analyze the two loop-contributions and deduce four guiding rules which are used in Sec.~\ref{Sec:n-order}, in order to demonstrate that only ladder-like diagrams are responsible for $[\alpha_s^n \log^{2n}(\xi-x)]/(x-\xi)$ contributions, which we then compute and resum. We end up with conclusions in Sec.~\ref{Sec:conclusion}. Two appendices give technical details on the analysis of the pole positions entering the loop integrals, and on  integrals used to extract the dominant contributions.

\section{One-loop analysis based on Ward identities}
\label{Sec:1-order}

In this section we analyze the one-loop diagrams in details without making any approximation to understand which diagrams give contribution at order $[\alpha_s \, \log^2(\xi-x)]/(x-\xi)$ and which give less singular contributions in light-like gauge.
We explicitly show that the net contribution to $[\log^2(\xi-x)]/(x-\xi)$ terms arises from the box-diagram in the case of cutting the gluonic line. Moreover, this analysis precisely identifies the part of the phase space that is responsible for this contribution.
\subsection{Self energy}
\label{SubSec:Self_Energy}

Let us start with the self-energy diagram, illustrated in Fig.~\ref{Fig:self-energy}. 
\def\factor{.9}
\begin{figure}
\psfrag{k3}{$S$}

\psfrag{kg1}{\raisebox{.2cm}{\hspace{-1.1cm}\scalebox{\factor}{$p_1 - 2 \xi \, p_2$}}}
\psfrag{kg2}{\raisebox{.2cm}{\hspace{.2cm}\scalebox{\factor}{$p_1$}}}

\psfrag{k1}{\raisebox{.2cm}{\hspace{.2cm}\scalebox{\factor}{$k$}}}

\psfrag{k4}{}
\psfrag{kl}{\scalebox{\factor}{$\hspace{-.7cm}p_1+(x-\xi) p_2$}}
\psfrag{kr}{\!\!\!\scalebox{\factor}{$\hspace{-.7cm}p_1+(x-\xi) p_2$}}

\psfrag{ki}[L]{\hspace{-.9cm}\scalebox{\factor}{$(x+\xi)p_2$}}
\psfrag{kf}[R]{\hspace{1.1cm} \scalebox{\factor}{$(x-\xi)p_2$}}

\centerline{\scalebox{.9}{\includegraphics[width=10cm]{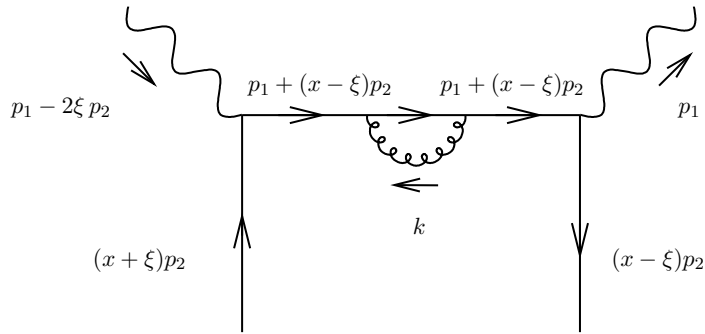}}}
\caption{One-loop self energy diagram.}
\label{Fig:self-energy}
\end{figure}
The numerator reads
\bea
({\rm Num})_{\rm S.E.}&=&\tr
\bigg\{\ps_2\g^{\sigma}_{\perp}\big[\ps_1+(x-\xi)\ps_2\big]\g^{\nu}\big[\ps_1+(x-\xi)\ps_2-\ks\big]\g^{\mu}\big[\ps_1+(x-\xi)\ps_2\big]\g_{\perp\sigma}\bigg\}\nonumber \\
&& \hspace{5cm}\times \bigg \{g_{\mu\nu}-\frac{k_\mu p_{1\nu}+k_\nu p_{1\mu}}{k\cdot p_1}\bigg\} \ .
\eea
After some algebra, one realizes that the gauge part of the numerator vanishes
\beq
({\rm Num})_{\rm gauge}=0 \ . 
\eeq
Since this is the case, the self energy diagram is exactly the same in Feynman and in light-like gauge. This diagram is calculated in Feynman gauge in Ref.~\cite{Pire:2011st} and it is shown that only single $\log$'s arise. Hence, the self-energy diagram does not contribute to $\frac{\log^2(\xi-x)}{(x-\xi)}$ terms.
\subsection{Right vertex, left vertex and box diagram}
\label{SubSec:All_Together}


\begin{figure}
\psfrag{k3}{\scalebox{\factor}{$\hspace{-1.4cm} p_1+(x-\xi) \,p_2$}}

\psfrag{kg1}{\raisebox{.4cm}{\hspace{-.5cm}\scalebox{\factor}{$p_1 - 2 \xi \, p_2$}}}
\psfrag{kg2}{\raisebox{.4cm}{\hspace{.8cm}\scalebox{\factor}{$p_1$}}}

\psfrag{k1}{\raisebox{.5cm}{\hspace{.2cm}\scalebox{\factor}{$k$}}}

\psfrag{k4}{\scalebox{\factor}{$\hspace{-1.3cm} p_1+(x-\xi) \, p_2+k$}}

\psfrag{ki}[L]{\hspace{0.3cm}\scalebox{\factor}{$(x+\xi)p_2$}}
\psfrag{kf}[R]{\hspace{1.1cm} \scalebox{\factor}{$(x-\xi)p_2$}}

\psfrag{k2}[L]{\hspace{1.6cm}\scalebox{\factor}{$(x-\xi)p_2+k$}}
\centerline{\scalebox{.8}{\includegraphics[width=11cm]{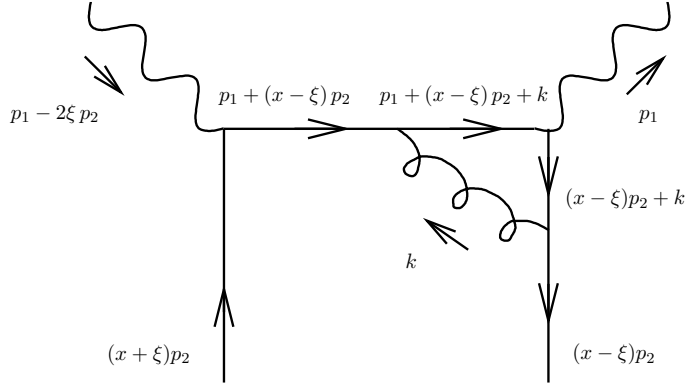}}}
\caption{Right-vertex diagram.}
\label{Fig:right-vertex}
\end{figure}

\begin{figure}
\psfrag{k3}{\hspace{-1.3cm}$k+p_1+(x-\xi) \, p_2$}
\psfrag{kg1}{\raisebox{.2cm}{\hspace{-1.1cm}$p_1 - 2 \xi \, p_2$}}
\psfrag{kg2}{\raisebox{.2cm}{\hspace{.2cm}$p_1$}}
\psfrag{k2}{$\hspace{-1.8cm}k+(x+\xi) \, p_2$}
\psfrag{k4}{$k+(x-\xi) \, p_2$}
\psfrag{k1}{$\ k$}
\psfrag{ki}[L]{\hspace{-.4cm}$(x+\xi)p_2$}
\psfrag{kf}[R]{\hspace{1.1cm} $(x-\xi)p_2$}
\centerline{\includegraphics[width=7cm]{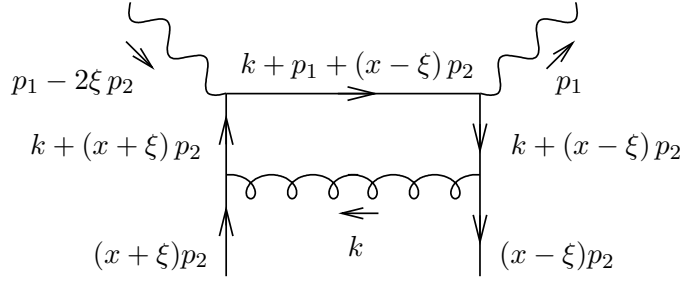}}
\caption{The one-loop box diagram.}
\label{Fig:1-loop-box}
\end{figure}

The numerator for the right-vertex diagram shown in Fig.~\ref{Fig:right-vertex} in the light-like gauge is written as
\bea
({\rm Num})_{\rm R.V.}=\tr\bigg\{\ps_2\g^{\mu}\big[\ks+(x-\xi)\ps_2\big]\g^{\sigma}_{\perp}\big[\ks+\ps_1+(x-\xi)\ps_2\big]\g^{\nu}\big[\ps_1+(x-\xi)\ps_2\big]\g_{\perp\sigma}\bigg\}\nonumber \\
\hspace{4cm} \times \bigg\{g_{\mu\nu}-\frac{k_\mu p_{1\nu}+k_\nu p_{1\mu}}{k\cdot p_1}\bigg\} \, .\
\label{num-right-vertex}
\eea
After some algebra, one gets
\bea
\label{num-right-vertex-gmunu}
({\rm Num})_{\rm R.V. \,g_{\mu \nu}}&=&-8s^2(1+\alpha)(\beta+x-\xi) \ , \\
\label{num-right-vertex-gauge2}
({\rm Num})_{\rm R.V. \,gauge}&=&8s(\beta+x-\xi)\bigg[s(1+\alpha)+\frac{k^2_{\perp}}{\beta}\bigg]\,.
\eea
Hence, the whole numerator reads
\beq
({\rm Num})_{\rm R.V.}=8s(\beta+x-\xi)\frac{k^2_{\perp}}{\beta}\,.
\eeq
Hereafter, $k_{\perp}$ (resp. $\kb$) denotes the transverse component of the gluon momentum in Minkowski (resp. Euclidean) space and we use $k^2_{\perp}=-\kb^2$ whenever it is needed. 
Hence, 
the integral for the right-vertex diagram is\footnote{The angular integration is straightforward for our calculation, which we emphasize through the notation $d_Nk=|k|^{N-1}d|k|$. This angular integration gives a factor of $(2\pi)^n$ for $n$-loops, from which $\pi^n$  is included inside the parenthesis of  Eq.~(\ref{def-Kn}) and $2^n$ is gathered inside $I_n.$}
\beq
I_{\rm R.V.}\!=\!-\frac{s}{2} \!\int \!d\alpha \, d\beta \, d_2\kb ~8s\frac{\kb^2}{\beta}(\beta+x-\xi)\frac{1}{s(x-\xi)}
\frac{1}{\big[k+(x-\xi)p_2)\big]^2}\frac{1}{k^2}\frac{1}{\big[k+p_1+(x-\xi)p_2)\big]^2}\,.
\label{I_RV}
\eeq
Similarly, the numerator for the left-vertex diagram in light-like gauge is written as
\bea
({\rm Num})_{\rm L.V.}=\tr\bigg\{\ps_2\g_{\perp}^{\sigma}\big[\ps_1+(x-\xi)\ps_2\big]\g^{\mu}\big[\ks+\ps_1+(x-\xi)\ps_2\big]\g_{\perp \sigma}\big[\ks+(x+\xi)\ps_2\big]\g^{\nu}\bigg\}
\nonumber \\
\hspace{3.5cm}\times \bigg\{g_{\mu\nu}-\frac{k_\mu p_{1\nu}+k_\nu p_{1\mu}}{k\cdot p_1}\bigg\} \ . \, \,
\eea
After some algebra one gets 
\bea
({\rm Num})_{{\rm L.V.} \,g_{\mu\nu}}&=&-8s^2(\beta+x+\xi)(1+\alpha)\ ,\\
({\rm Num})_{\rm L.V. \, gauge}&=&\frac{4}{\beta}(\beta+x+\xi)\bigg[2s^2\beta(1+\alpha)+2sk_{\perp}^2\bigg] \ .
\eea
Hence, the whole numerator reads
\beq
({\rm Num})_{\rm L.V.}=8s\frac{k_{\perp}^2}{\beta}(\beta+x+\xi)\ .
\eeq
One can write the integral for the left-vertex as
\beq
I_{\rm L.V.}\!=\!-\frac{s}{2}\!\!\int \!\!d\alpha \, d\beta \,d_2\kb\! ~\bigg[\!8s\frac{\kb^2}{\beta}(\beta+x+\xi)\frac{1}{s(x-\xi)}\!\bigg]\!
\frac{1}{\big[k+(x+\xi)p_2)\big]^2}\frac{1}{k^2}\frac{1}{\big[k+p_1+(x-\xi)p_2)\big]^2}.
\eeq
Let us now calculate the numerator for the box diagram, illustrated in Fig.~\ref{Fig:1-loop-box}, which is 
\bea
&&({\rm Num})_{\rm box}=\tr\bigg\{\ps_2\g^{\nu}\big[\ks+(x-\xi)\ps_2\big]\g^{\sigma}_{\perp}\big[\ks+\ps_1+(x-\xi)\ps_2\big]\g_{\perp\sigma}\big[\ks+(x+\xi)\ps_2\big]\g^{\mu}\bigg\}
\nonumber \\
&&\hspace{8cm}\times\bigg\{g_{\mu\nu}-\frac{k_\mu p_{1\nu}+k_\nu p_{1\mu}}{k\cdot p_1}\bigg\} \ . \
\label{Num_box}
\eea
The $g_{\mu\nu}$ part of the numerator reads
\beq
({\rm Num})_{{\rm box}\,g_{\mu\nu}}=\tr\bigg\{\big[\ks+(x+\xi)\ps_2\big]\g^{\mu}\ps_2\g_{\mu}\big[\ks+(x-\xi)\ps_2\big]\g^{\sigma}_{\perp}\big[\ks+\ps_1+(x-\xi)\ps_2\big]\g_{\perp\sigma}\bigg\} \ .
\eeq
Using $\g^{\mu}\ps_2\g_{\mu}=-2\ps_2$, 
it can be written as
\beq
({\rm Num})_{{\rm box}\,g_{\mu\nu}}=-2 \, \tr\bigg\{\big[\ks+(x+\xi)\ps_2\big]\ps_2\big[\ks+(x-\xi)\ps_2\big]\g^{\sigma}_{\perp}\big[\ks+\ps_1+(x-\xi)\ps_2\big]\g_{\perp\sigma}\bigg\} \ .
\eeq
Applying the Ward identity by noting that $p_2$ inside the trace can be put in the form
\beq
\label{ps_2 rewrite}
p^{\mu}_2=\frac{1}{2\xi}\bigg(\big[k+(x+\xi)p_2\big]-\big[k+(x-\xi)p_2\big]\bigg)^{\mu} \ ,
\eeq 
one gets
\bea
&&({\rm Num})_{{\rm box}\,g_{\mu\nu}}\!=\!-\frac{8}{\xi}\big[k+(x+\xi)p_2\big]^2 \! \bigg\{k^2_\perp-(\beta+x-\xi)\frac{s}{2}\bigg\}\nonumber\\
&&\hspace{4cm}+\frac{8}{\xi}\big[k+(x-\xi)p_2\big]^2\!\bigg\{k^2_\perp-(\beta+x+\xi)\frac{s}{2}+\xi\alpha s\bigg\}  .
\eea
In this way, we effectively reduce the box diagram to right and left vertex diagrams.
%
The gauge part of the numerator is written as
\bea
({\rm Num})_{{\rm box}\,gauge}&=&-\frac{2}{\beta s}\tr\bigg\{\big[\ks+(x+\xi)\ps_2\big]\ps_1\ps_2\ks\big[\ks+(x-\xi)\ps_2\big]\g^{\sigma}_{\perp}\big[\ks+\ps_1+(x-\xi)\ps_2\big]\g_{\perp\sigma}\bigg\} \nonumber \\
&&\hspace{-1.2cm}-\frac{2}{\beta s}\tr\bigg\{\big[\ks+(x+\xi)\ps_2\big]\ks\ps_2\ps_1\big[\ks+(x-\xi)\ps_2\big]\g^{\sigma}_{\perp}\big[\ks+\ps_1+(x-\xi)\ps_2\big]\g_{\perp\sigma}\bigg\}\,.
\eea
Since $p^2_2=0$, one can add or subtract $\ps_2$ from $\ks$ when they are appearing next to each other inside the trace without spoiling the result, which allows to cancel one of the fermionic propagators. 
The whole numerator for the box diagram is given as
\bea
({\rm Num})_{\rm box}&=&8\big[k+(x-\xi)p_2\big]^2\bigg\{\frac{1}{\xi}\bigg[k^2_{\perp}-(\beta+x+\xi)\frac{s}{2}+\xi\alpha s\bigg]+\frac{s}{\beta}(1+\alpha)(\beta+x+\xi)\bigg\} \nonumber \\
&-&8\big[k+(x+\xi)p_2\big]^2\bigg\{\frac{1}{\xi}\bigg[k^2_{\perp}-(\beta+x-\xi)\frac{s}{2}\bigg]-\frac{s}{\beta}(1+\alpha)(\beta+x-\xi)\bigg\}\,.
\eea
Rewriting the transverse component of the gluon momentum in Euclidean space,  the integral for the box diagram is 
\bea
\label{boxint-1}
I_{\rm box}&=&-\frac{s}{2}\int d\alpha \,d\beta\, d_2\kb~8\bigg\{\frac{1}{\xi}\bigg[\kb^2+(\beta+x+\xi)\frac{s}{2}-\xi\alpha s\bigg]-\frac{s}{\beta}(1+\alpha)(\beta+x+\xi)\bigg\}\nonumber \\
&&\hspace{4cm}\times\frac{1}{k^2}\frac{1}{\big[k+(x+\xi)p_2\big]^2}\frac{1}{\big[k+p_1+(x-\xi)p_2\big]^2}\\
&&-\frac{s}{2}\int d\alpha \,d\beta\, d_2\kb~(-8)\bigg\{\frac{1}{\xi}\bigg[\kb^2+(\beta+x-\xi)\frac{s}{2}\bigg]+\frac{s}{\beta}(1+\alpha)(\beta+x-\xi)\bigg\}\nonumber \\
&&\hspace{4cm}\times\frac{1}{k^2}\frac{1}{\big[k+p_1+(x-\xi)p_2\big]^2}\frac{1}{\big[k+(x-\xi)p_2\big]^2} \,.
\label{boxint-2}
\eea
Note that  the term (\ref{boxint-1}) is effectively the same as the left-vertex diagram since the fermionic propagator on the outgoing quark line is cancelled. On the other hand,  the term (\ref{boxint-2}) is effectively the same as the right-vertex diagram since the fermionic propagator on the incoming quark line is cancelled. This decomposition can be symbolically illustrated as
\beq
\psfrag{k}{}
\psfrag{k1}{}
\psfrag{k2}{}
\psfrag{k3}{}
\psfrag{k4}{}
\psfrag{R1}{}
\psfrag{R2}{}
\psfrag{L1}{}
\psfrag{L2}{}
\psfrag{ki}{}
\psfrag{kf}{}
\psfrag{kg1}{}
\psfrag{kg2}{}
\raisebox{-.5cm}{\includegraphics[width=3cm]{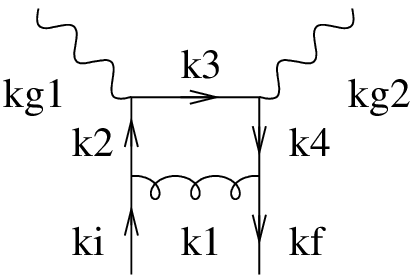}} = 
\raisebox{-.5cm}{\includegraphics[width=2.95cm]{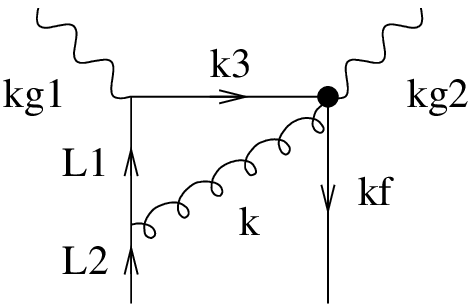}}+\raisebox{-.5cm}{\includegraphics[width=3.1cm]{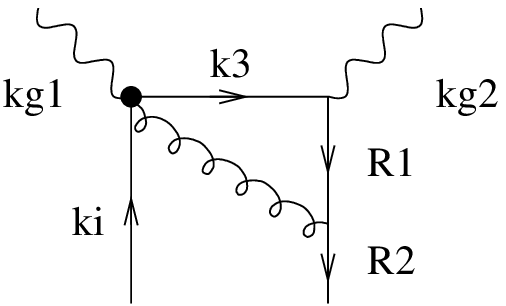}}\,.
\label{symbol-box-expansion}
\eq
Hence, we can write the integral as a sum of box, left-vertex and right-vertex diagrams in the following way
\beq
\label{box+LV+RV}
I_{\rm box\,+ \, L.V.\, +\, R.V.}=I_{\rm E.L.V.}+I_{{\rm E.R.V.}}\,,
\eeq
where
\bea
\label{ELV}
I_{\rm E.L.V.}&=&-\frac{s}{2}\int d\alpha \,d\beta\, d_2\kb~8\bigg\{\frac{1}{\xi}\bigg[\kb^2+(\beta+x+\xi)\frac{s}{2}-\xi\alpha s\bigg]-\frac{s}{\beta}(1+\alpha)(\beta+x+\xi)\nonumber \\
&&\hspace{1cm}+\frac{\kb^2}{\beta}\frac{(\beta+x+\xi)}{(x-\xi)}\bigg\}\frac{1}{k^2}\frac{1}{\big[k+(x+\xi)p_2\big]^2}\frac{1}{\big[k+p_1+(x-\xi)p_2\big]^2}\,, 
\eea
which symbolically means that
\beq
\psfrag{k}{}
\psfrag{k1}{}
\psfrag{k2}{}
\psfrag{k3}{}
\psfrag{k4}{}
\psfrag{R1}{}
\psfrag{R2}{}
\psfrag{L1}{}
\psfrag{L2}{}
\psfrag{ki}{}
\psfrag{kf}{}
\psfrag{kg1}{}
\psfrag{kg2}{}
 I_{\rm E.L.V.}= 
\raisebox{-.5cm}{\includegraphics[width=3cm]{one-loop-left-effective-no-arrows.eps}}+\raisebox{-.5cm}{\includegraphics[width=2.35cm]{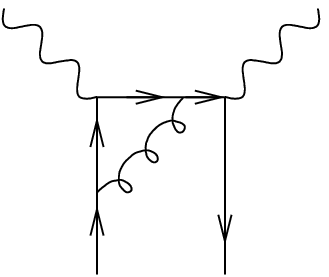}}\,,
\label{symbol-def-ELV}
\eq
while
\bea
\label{ERV}
I_{{\rm E.R.V.}}&=&-\frac{s}{2}\int d\alpha \, d\beta \, d_2\kb~(-8)\bigg\{\frac{1}{\xi}\bigg[\kb^2+(\beta+x-\xi)\frac{s}{2}\bigg]+\frac{s}{\beta}(1+\alpha)(\beta+x-\xi)\nonumber \\
&&\hspace{1cm} -\frac{\kb^2}{\beta}\frac{(\beta+x-\xi)}{(x-\xi)}\bigg\}\frac{1}{k^2}\frac{1}{\big[k+p_1+(x-\xi)p_2\big]^2}\frac{1}{\big[k+(x-\xi)p_2\big]^2} \,,
\eea
which symbolically means that
\beq
\psfrag{k}{}
\psfrag{k1}{}
\psfrag{k2}{}
\psfrag{k3}{}
\psfrag{k4}{}
\psfrag{R1}{}
\psfrag{R2}{}
\psfrag{L1}{}
\psfrag{L2}{}
\psfrag{ki}{}
\psfrag{kf}{}
\psfrag{kg1}{}
\psfrag{kg2}{}
 I_{\rm E.R.V.}= 
\raisebox{-.5cm}{\includegraphics[width=3cm]{one-loop-right-effective-no-arrows.eps}}+\raisebox{-.5cm}{\includegraphics[width=2.21cm]{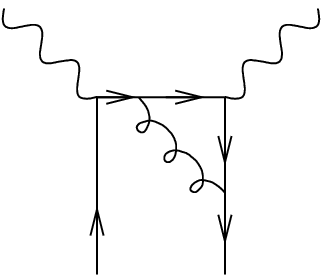}}\,.
\label{symbol-def-ERV}
\eq
We refer to  Eq.~(\ref{ELV}) as effective left-vertex (E.L.V.) and  Eq.~(\ref{ERV}) as effective right-vertex (E.R.V.) for simplicity and we consider the integrals separately.

We start our analysis with E.L.V. and use Cauchy integration to integrate over $\alpha$. A detailed analysis for the distribution of the poles is given in App.~\ref{Ap:beta-range}. We are free to choose to close either on the two  poles corresponding to cutting the gluonic line, i.e. $\alpha_g=\frac{\kb^2}{s\beta}$ and left fermionic line, i.e. $\alpha_f=\frac{\kb^2}{s(\beta+x+\xi)}$, or
on the right and $s-$channel fermionic line, which we avoid.
 The integration over $\kb$ is performed by using dimensional regularization. Then, the integral for E.L.V. reads
\beq
\label{genralint}
I_{\rm E.L.V.}=-2\pi i\bigg[\int_0^{\xi-x}d\beta\int_0^{\infty}d_N\kb~Res_{\alpha_g}+\int_{-\xi-x}^{\xi-x}d\beta\int_0^{\infty}d_N\kb~Res_{\alpha_f}\bigg]\,,
\eeq
where 
\bea
\label{resalphag}
Res_{\alpha_g}&\!=&\!-4\frac{1}{(x-\xi)}\bigg[\frac{\beta}{\xi(x+\xi)}-\frac{1}{(x+\xi)}+\frac{(\beta+x+\xi)}{(x+\xi)(x-\xi)}-\frac{(\beta+x+\xi)}{2\xi(\beta+x-\xi)}\bigg]\frac{1}{\kb^2+\frac{\beta(\beta+x-\xi)s}{(x-\xi)}}\nonumber \\
&&-4\frac{1}{(x-\xi)}\frac{(\beta+x+\xi)}{(x+\xi)}\bigg[\frac{\beta}{2\xi}-1\bigg]\frac{(x-\xi)}{\beta(\beta+x-\xi)}\frac{1}{\kb^2}\,,
\eea
and
\bea
\label{resalphaf}
Res_{\alpha_f}&=&-4\frac{1}{(x+\xi)2\xi}\bigg\{(\beta+x+\xi)\bigg[\frac{1}{\xi}+\frac{1}{x-\xi}+\frac{(x+\xi)}{(x-\xi)}\frac{1}{(\beta+x-\xi)}\bigg]-1\bigg\} \nonumber \\
&&\times \frac{1}{\kb^2-\frac{(\beta+x+\xi)(\beta+x-\xi)s}{2\xi}} 
-4\frac{1}{(x+\xi)}\frac{(\beta+x+\xi)}{(\beta+x-\xi)}\bigg[\frac{1}{\beta}-\frac{1}{2\xi}\bigg]\frac{1}{\kb^2}\,,
\eea
with $N=2-\epsilon_{UV}=2+\epsilon_{IR}$.
In order to get the above expressions we have used the following relation
\beq
\label{identity1}
\frac{1}{A\,B}=\bigg[\frac{1}{A}-\frac{1}{B}\bigg]\frac{1}{B-A} \ .
\eeq
Using the fact that in dimensional regularization any integral without scale vanishes, the second line of  Eq.~(\ref{resalphag}) and the last term of Eq.~(\ref{resalphaf}) give zero. The ultraviolet divergence in $\kb$ integral in both expressions is taken into account by renormalization and we are only interested in the finite part.

Keeping all these remarks in mind, let us first calculate the gluonic pole contribution, 
\beq
\psfrag{k2}{}
\psfrag{k3}{}
\psfrag{k4}{}
\psfrag{R1}{}
\psfrag{R2}{}
\psfrag{L1}{}
\psfrag{L2}{}
\psfrag{ki}{}
\psfrag{kf}{}
\psfrag{kg1}{}
\psfrag{kg2}{}
 I_{\rm E.L.V., \, g}= 
\raisebox{-.5cm}{
\psfrag{k}{\raisebox{.32cm}{\hspace{-.32cm}\rotatebox{120}{\Large $-$}}}
\includegraphics[width=3cm]{one-loop-left-effective-no-arrows.eps}}+
\raisebox{-.5cm}{\psfrag{k}{\raisebox{.22cm}{\hspace{-.4cm}\rotatebox{120}{\Large $-$}}}
\includegraphics[width=3cm]{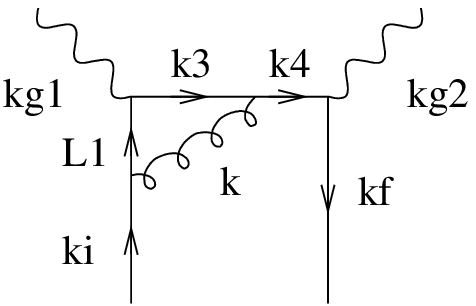}}\,.
\label{symbol-def-ELV-cut-gluon}
\eq
The integration over $\kb$ in the first term of  Eq.~(\ref{genralint}) and using Eq.~(\ref{resalphag}) gives 
\bea
\label{ELVg}
&&I_{\rm E.L.V.,\,g}=4\frac{2\pi i}{x-\xi}\int_0^{\xi-x}d\beta \bigg[\frac{\beta}{\xi(x+\xi)}-\frac{1}{(x+\xi)}+\frac{(\beta+x+\xi)}{(x+\xi)(x-\xi)}-\frac{(\beta+x+\xi)}{2\xi(\beta+x-\xi)}\bigg] \nonumber \\
&&\hspace{8cm}\times \Gamma(\epsilon_{UV}) \bigg[\frac{s\beta(\beta+x-\xi)}{x-\xi}\bigg]^{\epsilon_{IR}} \, .
\eea
Before integrating over $\beta$, in order to simplify the calculation, one should stress that we are looking for the terms that contribute to the $\frac{\log^2(\xi-x)}{(x-\xi)}$ terms, i.e. the most singular terms. This corresponds to terms that are most singular at the limits of the integration. Hence, for the $I_{\rm E.L.V.,\,g}$  integral we are interested in $\frac{1}{\beta}$ terms which are singular at the lower limit, and $\frac{1}{\beta+x-\xi}$ terms which are singular at the upper limit. In Eq.~(\ref{ELVg}) there is no term that is proportional to $\frac{1}{\beta}$. Thus, there is no contribution from $\frac{1}{\beta}$ in  $I_{\rm E.L.V.,\,g}$. The second singularity that should be considered is $\frac{1}{\beta+x-\xi}$  and for this type of singularity the integral reads
\beq
I_{\rm E.L.V.,\,g}=-4\frac{2\pi i}{x-\xi}\Gamma(\epsilon_{UV})\int_0^{\xi-x}d\beta (\beta+x-\xi)^{\epsilon_{IR}-1}\,.
\eeq
Integration over $\beta$ is straightforward after this point. Hence, integrating over $\beta$ and expanding the expression in the limit $\epsilon_{IR,\,UV}\rightarrow 0$, the finite part reads
\beq
\label{net1loop}
I^{\rm fin.}_{\rm E.L.V.,\,g}=-4\frac{2\pi i}{x-\xi}\frac{1}{2!}\log^2(\xi-x)\,.
\eeq
Let us emphasize that this contribution is coming only from the last term of the first line in Eq.~(\ref{resalphag}) which is originated from the box diagram. 

In the same way as in the calculation of the gluonic pole, the contribution from the fermionic pole, 
\beq
\psfrag{k}{}
\psfrag{k2}{}
\psfrag{k3}{}
\psfrag{k4}{}
\psfrag{R1}{}
\psfrag{R2}{}
\psfrag{L2}{}
\psfrag{ki}{}
\psfrag{kf}{}
\psfrag{kg1}{}
\psfrag{kg2}{}
 I_{\rm E.L.V.,\,f}= 
\raisebox{-.5cm}{
\psfrag{L1}{\raisebox{.15cm}{\Large$\hspace{.22cm}-$}}
\includegraphics[width=3cm]{one-loop-left-effective-no-arrows.eps}}+
\raisebox{-.5cm}{
\psfrag{L1}{\raisebox{.15cm}{\Large$\hspace{.22cm}-$}}
\includegraphics[width=3cm]{one-loop-leftLARGE-BISno-arrows.eps}}\,,
\label{symbol-def-ELV-cut-fermion}
\eq
after $\kb$ integration, gives
\bea
I_{\rm  E.L.V.,\,f}&\!=&\!4\frac{2\pi i}{(x+\xi)2\xi}\int_{-\xi-x}^{\xi-x}\!\!\!d\beta\bigg\{(\beta+x+\xi)\bigg[\frac{1}{\xi}+\frac{1}{x-\xi}+\frac{(x+\xi)}{(x-\xi)}\frac{1}{(\beta+x-\xi)}\bigg]-1\bigg\}\Gamma(\epsilon_{UV})\nonumber \\
&&\hspace{6cm}\times\bigg[\frac{s(\beta+x+\xi)(\beta+x-\xi)}{2\xi}\bigg]^{\epsilon_{IR}}\,.\
\eea
The most singular terms that we are looking for in this integration are $\frac{1}{\beta+x+\xi}$ and $\frac{1}{\beta+x-\xi}$ terms which are singular at the lower and upper limit respectively. It is obvious from the expression that there are no $\frac{1}{\beta+x+\xi}$ type singularity. For $\frac{1}{\beta+x-\xi}$ type singularity, the integral reads
\beq
I_{\rm E.L.V.,\,f}=4\frac{2\pi i}{x-\xi}\Gamma(\epsilon_{UV})\int_{-x-\xi}^{\xi-x}d\beta (\beta+x-\xi)^{\epsilon_{IR}-1}\,.
\eeq
Again, integrating over $\beta$ and expanding the result in the limit  $\epsilon_{IR,\, UV}\rightarrow 0$, the finite part reads
\beq
I^{\rm fin.}_{\rm E.L.V.,\,f}= 4\frac{2\pi i}{x-\xi}\frac{1}{2!}\log^2(2\xi) \ ,
\eeq
which is less singular than Eq.~(\ref{net1loop}).
 Note that taken separately, each of these two diagrams which involve cutting the fermionic line lead to  $\log^2(\xi-x)/(x-\xi)$ terms, but this type of contributions add to zero at the end.

A similar analysis can be made for the effective right vertex, E.R.V.  Again two poles corresponding to cutting the gluonic line, i.e. $\alpha_g=\frac{\kb^2}{s\beta}$ and right fermionic line, i.e. $\alpha_f=\frac{\kb^2}{s(\beta+x-\xi)}\,,$ are considered. By using dimensional regularization the integral for E.R.V. is written as
\beq
I_{{\rm E.R.V.}}=-2\pi i\bigg[\int_0^{\xi-x}d\beta \int_0^{\infty}d_N\kb~Res_{\alpha_g}+\int_{-\xi-x}^{\xi-x}d\beta \int_0^{\infty}d_N\kb~Res_{\alpha_f}\bigg]\,,
\eeq
where 
\bea
\label{resalphagR}
Res_{\alpha_g}&=&4\frac{1}{(x-\xi)^2}\bigg[\frac{\beta}{\xi}+1-\frac{(\beta+x-\xi)}{x-\xi}-\frac{(x-\xi)}{2\xi}\bigg]\frac{1}{\kb^2+\frac{\beta(\beta+x+\xi)s}{x-\xi}} \nonumber \\
&&+4\frac{1}{(x-\xi)}\bigg[\frac{1}{2\xi}+\frac{1}{\beta}\bigg]\frac{1}{\kb^2}\,,
\eea
and
\beq
\label{resalphafR}
Res_{\alpha_f}=-4\frac{1}{(x-\xi)}\bigg\{\bigg[\frac{1}{\xi(\beta+x-\xi)}+\frac{1}{\beta(\beta+x-\xi)}-\frac{1}{\beta(x-\xi)}\bigg]+s\bigg(\frac{1}{2\xi}+\frac{1}{\beta}\bigg)\frac{1}{\kb^2}\bigg\}\,,
\eeq
with as above $N=2-\epsilon_{UV}=2+\epsilon_{IR}$.

Again, using the fact that any scaleless integral vanishes in dimensional regularization, one  can immediately set the second line of Eq.~(\ref{resalphagR}) to zero. Moreover, one can  see that the finite part of  Eq.~(\ref{resalphafR}) vanishes totally since one of the terms is $\kb$ independent and the second is scaleless. Hence, the only non-vanishing part of $I_{{\rm ERV}}$ after integrating over $\kb$ reads
\beq
\label{ERVafterk}
I_{{\rm E.R.V.}}=-4\frac{2\pi i}{(x-\xi)^2}\int_0^{\xi-x}d \beta \bigg[\frac{\beta}{\xi}+1-\frac{(\beta+x-\xi)}{x-\xi}-\frac{(x-\xi)}{2\xi}\bigg] \Gamma(\epsilon_{UV})\bigg[\frac{s(\beta+x-\xi)}{x-\xi}\bigg]^{\epsilon_{IR}}\,.
\eeq
The most singular terms that may contribute to $\frac{\log^2(\xi-x)}{x-\xi}$ terms are $\frac{1}{\beta}$ and $\frac{1}{\beta+x-\xi}$ terms none of which are present in Eq.~(\ref{ERVafterk}). The most singular term in the $x\rightarrow\xi$ limit is $\frac{\log^2(\xi-x)}{x-\xi}$, hence $I_{{\rm E.R.V.}}$ does not give contribution to $\frac{\log^2(\xi-x)}{x-\xi}$ terms.

Thus, the above one-loop analysis shows that the only contribution to $\frac{\log^2(\xi-x)}{x-\xi}$ terms come from Eq.~(\ref{net1loop}) which is originated from the box diagram in the case of cutting the gluonic line around $\beta+x-\xi\approx 0$ in the phase space.
We have thus shown that
\beq
\psfrag{k2}{}
\psfrag{k3}{}
\psfrag{k4}{}
\psfrag{R1}{}
\psfrag{R2}{}
\psfrag{L1}{}
\psfrag{L2}{}
\psfrag{ki}{}
\psfrag{kf}{}
\psfrag{kg1}{}
\psfrag{kg2}{}
I_{\rm one \, loop}^{\rm dominant}= 
\raisebox{-.5cm}{
\psfrag{k}{\raisebox{.32cm}{\hspace{-.32cm}\rotatebox{120}{\Large $-$}}}
\includegraphics[width=3cm]{one-loop-left-effective-no-arrows.eps}}=-4\frac{2\pi i}{x-\xi}\frac{1}{2!}\log^2(\xi-x)\,.
\label{symbol-net-result-ELV-cut-gluon}
\eq

We would like to emphasize that the precision of our calculation does not permit us to fix the multiplicative coefficient $a$ of $(\xi-x)$ under logarithm, i.e. Eq.~(\ref{net1loop}) can be equivalently written as
\beq
\label{net1loopa}
I_{\rm one \, loop}^{\rm dominant}\approx-4\frac{2\pi i}{x-\xi}\frac{1}{2!}\log^2[a(\xi-x)] \ .
\eeq
The coefficient $a$ is fixed to $\frac{1}{2\xi}$ by comparing the form of the $\log^2(\xi-x)$ terms in the exact one-loop result  Eq.~(\ref{C1}). Moreover, the shift $\xi \rightarrow \xi-i\epsilon$ correctly takes into account the imaginary part of  Eq.~(\ref{C1}) leading to the following final formula
\beq
\label{net1loopiepsilon}
I_{\rm one \, loop}^{\rm dominant}\approx-4\frac{2\pi i}{x-\xi+i\epsilon}\frac{1}{2!}\log^2\bigg[\frac{\xi-x}{2\xi}-i\epsilon\bigg] \ .
\eeq
This matching condition will be repeatedly used in calculations in higher loops and also in the resummation process.

To conclude this section, we can thus state the first rule:
\\

\no
{\it (i) To extract the dominant behavior of the amplitude, it is sufficient to restrict ourselves to the contribution of the gluonic pole.}
\\

This rule will be extended later after studying in detail the two-loop contributions.

\section{One-loop in semi-eikonal approximation}
\label{Sec:1-order-ladder}

\begin{figure}[h!]
\psfrag{k3}{$S$}

\psfrag{kg1}{\raisebox{.2cm}{\hspace{-1.1cm}$p_1 - 2 \xi \, p_2$}}
\psfrag{kg2}{\raisebox{.2cm}{\hspace{.2cm}$p_1$}}
\psfrag{k2}{$L_1$}
\psfrag{k4}{$R_1$}
\psfrag{k1}{$k_1$}

\psfrag{ki}[L]{\hspace{-.9cm}$(x+\xi)p_2$}
\psfrag{kf}[R]{\hspace{1.1cm} $(x-\xi)p_2$}

\centerline{\includegraphics[width=5cm]{one-loop-large-direct.eps} \hspace{1.5cm}
\psfrag{k3}{}
\psfrag{kg1}{}
\psfrag{kg2}{}
\psfrag{k2}{}
\psfrag{k3}{}
\psfrag{k4}{}
\psfrag{k1}{}
\psfrag{ki}{}
\psfrag{kf}{}
 \includegraphics[width=5cm]{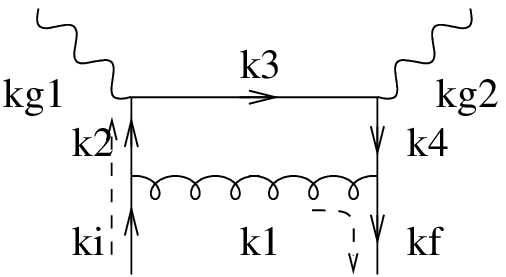}}
\caption{The one-loop ladder diagram which contribute in the light-like gauge to the leading  $[\alpha_s \ln^{2}(\xi-x)]/(x-\xi)$ terms in the perturbative expansion of the    DVCS  amplitude. The $p_2$ and $\perp$ momentum components are indicated. On the right, the dashed lines show the dominant momentum flows along the $p_2$ direction.} 
\label{Fig:1-loop}
\end{figure}

As explained in details in Sec.~\ref{Sec:1-order}, the dominant contribution for $x \to \xi$ is obtained from ladder-type diagrams at one-loop level. Thus, we
now concentrate on the box diagram, see Fig.~\ref{Fig:1-loop}, and we show that the result that was obtained without making any approximation can be reproduced by using eikonal techniques applied to the left fermionic line of the box diagram. We do not rely on the Ward identities of the previous section, but rather
expand the gluon propagator in the light-like gauge, using the fact that it will be considered to be on-shell. This will simplify much the analysis, similarly as in the case of   small-x physics considered in Ref.~\cite{Shuvaev:2006br}.
%
The corresponding integral $I_1$ reads
\beq
I_1=\frac{s}{2}\int d\alpha_1 \, d\beta_1 \, d_2\kb_1~({\rm Num})_1\frac{1}{L^2_1}\frac{1}{S^2}\frac{1}{R^2_1}\frac{1}{k^2_1}\,,
\eeq
where 
\begin{equation}
({\rm Num})_1=\tr\big\{ \not{p_2}\gamma_{\mu}[\not{k_1}+(x-\xi)\not{p_2}]\theta[\not{k_1}+(x+\xi)\not{p_2}]\gamma_{\nu} \big\}d^{\mu\nu}\,,
\end{equation}
\beq
L_1^2=\big[k_1+(x+\xi)p_2\big]^2 \  \  \ , \  \   S^2=\big[k_1+p_1+(x-\xi)p_2\big]^2 \  \  \ , \  \  R_1^2=\big[k_1+(x-\xi)p_2\big]^2\,,
\eeq
and  $\theta = \g^{\sigma}_{\perp}[\ks_1+\ps_1+(x-\xi)\ps_2]\g_{\sigma \perp} $. 

We now use an eikonal coupling for the left  quark line, and we treat the gluon  as soft with respect to this quark. Thus, in the incoming quark numerator $k_1+(x+\xi)p_2$  is replaced by $(x+\xi)p_2$.  Furthermore, we treat  this gluon  as soft with respect to the $s-$channel fermionic line, thus working in the limit $\alpha_1 \ll 1.$ 
This leads to the fact that $\theta$ can be approximated  as $\theta=-2\ps_1$. 
Note that this was checked in detail in Sec.~\ref{Sec:1-order}, where we have seen  that the dominant integration region corresponds indeed to the approximation  $\alpha_1 \ll 1$.


Since the gluon is on mass shell, the dominant contribution from the gluon propagator, $d^{\mu\nu}$, when written in terms of gluon polarization vectors, is given by 
\begin{equation}
d^{\mu\nu}\approx -\sum_{\lambda}\epsilon^{\mu}_{(\lambda)}\epsilon^{\nu}_{(\lambda)}\,.
\label{prop-lambda}
\end{equation}
The numerator with the eikonal coupling to left fermion leads to the expression 
\begin{equation}
({\rm Num})_1=-2(x+\xi)\sum_{\lambda}\tr\big\{ \not{p_2}\gamma_{\mu}[\not{k_1}+(x-\xi)\not{p_2}] \not{p_1}\not{p_2}\not{\epsilon}_{(\lambda)} \big\}(-\epsilon^{\mu}_{(\lambda)}) \ ,
\end{equation}
which, after using the Sudakov decomposition of gluon polarization vector in $p_1$ gauge,  
\begin{equation}
\epsilon^{\mu}_{(\lambda)}= \epsilon^{\mu}_{\perp (\lambda)} -2 \frac{\epsilon_{\perp (\lambda)}\cdot k_{\perp 1}}{\beta_1 s}p_1^{\mu}\,,
\end{equation} 
is rewritten as
\begin{equation}
\label{num1}
({\rm Num})_1=-2(x+\xi)\sum_{\lambda}\bigg(-2 \frac{\epsilon_{\perp (\lambda)}\cdot k_{\perp 1}}{\beta_1 s}\bigg)\tr\big\{ \not{p_2}\gamma_{\mu}[\not{k_1}+(x-\xi)\not{p_2}] \not{p_1}\not{p_2}\not{p_1}\big\}(-\epsilon^{\mu}_{(\lambda)})\,.
\end{equation}
Summing over the polarizations one gets
\begin{equation}
\label{sumoverpol}
\sum_{\lambda}\epsilon_{\perp (\lambda)}\cdot k_{\perp 1}\epsilon^{\mu}_{(\lambda)}=\bigg(-k^{\mu}_{\perp 1}+2\frac{k_{\perp 1}^2}{\beta_1 s}p^{\mu}_1\bigg)\,.
\end{equation}
Substituting Eq.~(\ref{sumoverpol}) to Eq.~(\ref{num1}), we arrive to the following expression for the numerator
\begin{eqnarray}
\label{numfinal}
({\rm Num})_1&=&-2(x+\xi)\bigg(\frac{-2}{\beta_1}\bigg)\tr\bigg\{\not{p_2}\bigg(\not{k_{\perp 1}}-2\frac{k_{\perp 1}^2}{\beta_1 s}\not{p_1}\bigg)[\not{k_1}+(x-\xi)\not{p_2}]\not{p_1}\bigg\} \nonumber \\
&=&-2(x+\xi)\bigg\{\frac{-2\kb_1^2}{\beta_1}\bigg[1+\frac{2(x-\xi)}{\beta_1}\bigg]\bigg\}2s \ .
\end{eqnarray}
The appearance of $\bigg[1+\frac{2(x-\xi)}{\beta_1}\bigg]$ in Eq.~(\ref{numfinal}) reflects the fact that the coupling of the right fermionic line is not the conventional eikonal coupling, since it takes into account some recoil effect.

The denominators with on shell gluon ($k^2=0$) are 
\begin{equation}
L_1^2=\alpha_1(x+\xi)s \  \  \  , \  \  \  R_1^2=-\kb_1^2+\alpha_1(\beta_1+x-\xi)s\  \  \  , \  \  \  S^2 = -\kb_1^2+(\beta_1+x-\xi)s \,.
\end{equation}
We use Cauchy integration to integrate over $\alpha_1$. The resulting expression  for the residue at the pole $\alpha_1=\frac{\kb_1^2}{s\beta_1}$  is
\begin{equation}
Res_{\alpha_1}=-4s\frac{1}{x-\xi}\bigg[1+\frac{2(x-\xi)}{\beta_1}\bigg]\frac{1}{\kb_1^2}\frac{1}{\kb_1^2-(\beta_1+x-\xi)s} \ ,
\end{equation}
which leads to $I_1$ integral (see App. ~\ref{Ap:beta-range} for the limits of the $\beta_1$ integral)
\begin{equation}
\label{integrall1}
I_1=4s\frac{2\pi i}{x-\xi}\int_0^{\xi-x}d\beta_1\int_0^{\infty}d_N\kb_1\bigg[1+\frac{2(x-\xi)}{\beta_1}\bigg]\frac{1}{\kb_1^2}\frac{1}{\kb_1^2-(\beta_1+x-\xi)s} \ .
\end{equation}
Using the identity (\ref{identity1}) for the last two propagators and taking into account the vanishing of the scaleless dimensionally regularized integrals, the expression (\ref{integrall1}) is 
\begin{equation}
I_1=-4\frac{2\pi i}{x-\xi}\int_0^{\xi-x}d\beta_1\int_0^{\infty}d_N\kb_1\frac{1}{(\beta_1+x-\xi)}\frac{1}{\kb_1^2-(\beta_1+x-\xi)s} \ ,
\end{equation}
in which also $\bigg[1+\frac{2(x-\xi)}{\beta_1}\bigg]$ is approximated by $-1$, in accordance with the ordering stated in  Eq.~(\ref{kinematics_beta}) and illustrated by Fig.~\ref{Fig:1-loop}.
This approximation is in accordance with the detailed analysis presented after Eq.~(\ref{ELVg}).
Integration over $\kb_1$ gives 
\begin{equation}
\label{regI1}
I_1=-4\frac{2\pi i}{x-\xi}\int_0^{\xi-x}d\beta_1\Gamma(\epsilon_{UV})\frac{1}{(\beta_1+x-\xi)}(\xi-x-\beta_1)^{\epsilon_{IR}} \ .
\end{equation}
 From the regularized expression (\ref{regI1}), one sees that in the $\beta_1$ integration the dominant contribution comes from the  region $\beta_1$ around 0. Keeping this in mind, the finite part of the integral (\ref{regI1}) can be determined from 
 \beq
I^{\rm fin.}_1=4\frac{2\pi i}{x-\xi}\int_0^{\cdots}d\beta_1\frac{1}{(\beta_1+x-\xi)}\log(\xi-x-\beta_1)\,,
 \eeq
in which we already remove the regularization.
 
Finally, after integrating over $\beta_1$ and using the matching  with the exact one-loop result, the dominant part of the one-loop diagram  is 
\beq
\psfrag{k3}{}
\psfrag{kg1}{}
\psfrag{kg2}{}
\psfrag{k2}{}
\psfrag{k4}{}
\psfrag{k1}{\raisebox{.4cm}{\hspace{-.05cm}\rotatebox{90}{\Large$-$}}}
\psfrag{ki}{}
\psfrag{kf}{}
I_{\rm one \, loop}^{\rm dominant} =\raisebox{-.8cm}{\includegraphics[width=4cm]{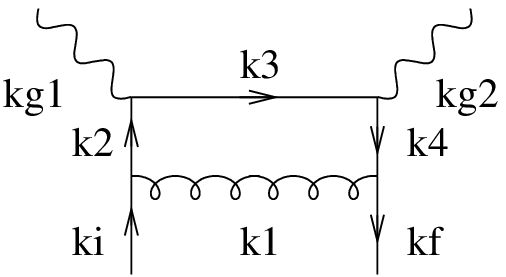}}
=-4\frac{2\pi i}{x-\xi+i\epsilon}\frac{1}{2!}\log^2\bigg[\frac{\xi-x}{2\xi}-i\epsilon\bigg]\,,
\label{box-cut-gluon}
\eq
in agreement with the result (\ref{net1loopiepsilon}).


\section{Two-loop order}
\label{Sec:2-order}

Let us examine the next order in the perturbative expansion. There are many diagrams contributing  but it can be shown that in the chosen gauge, the double box diagram dominates.



\subsection{Two-loop in semi-eikonal approximation}
\label{SubSec:ladder-2}

\begin{figure}[h!]
\psfrag{k3}{$S$}

\psfrag{kg1}{\raisebox{.2cm}{\hspace{-1.2cm}$p_1 - 2 \xi \, p_2$}}
\psfrag{kg2}{\raisebox{.2cm}{\hspace{.3cm}$p_1$}}
\psfrag{k2}{$\,k_2$}
\psfrag{k4}{$R_1$}
\psfrag{k1}{\,$k_1$}

\psfrag{ki}[L]{\hspace{-.9cm}$(x+\xi)p_2$}
\psfrag{kf}[R]{\hspace{1.1cm} $(x-\xi)p_2$}

\psfrag{L1}{$\hspace{-1.4cm}(x+\xi)\, p_2$}
\psfrag{L2}{$\!\!L_1$}
\psfrag{L3}{$\!\!L_2$}
\psfrag{R1}{$(x-\xi)\, p_2$}
\psfrag{R2}{$R_1$}
\psfrag{R3}{$R_2$}
\psfrag{delta}{$\ \ S$}
\centerline{\includegraphics[width=5.3cm]{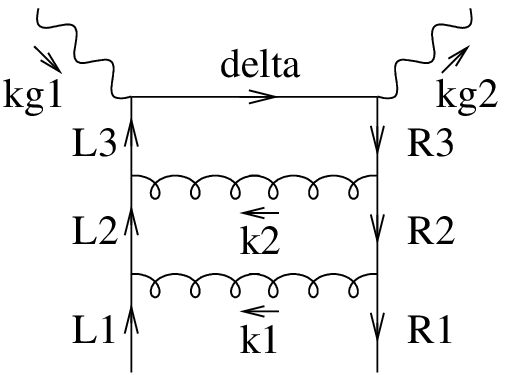} \hspace{1.5cm}
\psfrag{k3}{}
\psfrag{kg1}{}
\psfrag{kg2}{}
\psfrag{k2}{}
\psfrag{k3}{}
\psfrag{k4}{}
\psfrag{k1}{}
\psfrag{ki}{}
\psfrag{kf}{}
\psfrag{L1}{}
\psfrag{L2}{}
\psfrag{L3}{}
\psfrag{R1}{}
\psfrag{R2}{}
\psfrag{R3}{}
\psfrag{delta}{}
 \includegraphics[width=5.3cm]{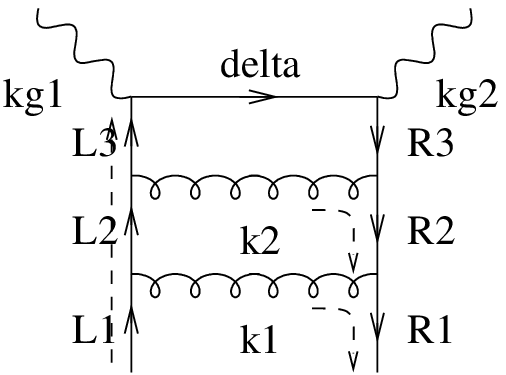}}
\caption{The two-loop ladder diagram which contribute in the light-like gauge to the leading  $[\alpha_s^2 \ln^{4}(\xi-x)]/(x-\xi)$ terms in the perturbative expansion of the    DVCS  amplitude. The $p_2$ and $\perp$ momentum components are indicated. On the right, the dashed line show the dominant momentum flows along the $p_2$ direction.}
\label{Fig:2-loop}
\end{figure}

The analysis for one-loop case showed that  the dominant contribution comes from the case where the gluon is on shell. So for the two-loop case we assume the same argument, i.e. both of the gluons are on shell. Moreover, assuming a strong ordering in $|\kb_i|$ and $\beta_i$ is natural,  since the singularities to be extracted are leading double logarithmic ones.
In practice, this means that we work in the approximation
\begin{equation}
|\kb_2| \gg |\kb_1|  \quad {\rm and} \quad 
x \sim \xi\gg \vert \beta_1\vert  \sim \vert x-\xi \vert \gg \vert x-\xi +\beta_1\vert \sim \vert \beta_2 \vert \,,
\end{equation}
which implies, for on-shell gluon (the fact that the gluon can be taken on-shell for the dominant contribution has been justified above), since we consider the gluons to be soft with respect to the $s$-channel fermion, that
\beq
1 \gg |\alpha_2| \gg |\alpha_1|\,.
\label {ordering-alpha-two-loop}
\eq
Note that the reverse ordering $|\kb_2| \ll |\kb_1|$ would lead to a suppressed contribution, due to a non maximal number of collinear 
singularities, which can be traced when evaluating the virtualities of the various loop momenta, as we will show in Sec.~\ref{SubSec:suppressed-2-loop}.

As a consequence of these two assumptions we arrive to the fact that the coupling to the left fermionic line can be considered as eikonal coupling whereas the coupling to the right fermionic line is beyond eikonal approximation in a way that it takes into account the recoil effect (see Eq. (\ref{numfinal}) and the following remark.) Then, one can write the integral $I_2$ for the two-loop ladder diagram as
\beq
\label{def:I2}
I_2=\bigg(\frac{s}{2}\bigg)^2\int d\alpha_1\,d\beta_1 \, d_2{\kb_1}\int  d\alpha_2 \,d\beta_2\,d_2{\kb_2}({\rm Num})_2\frac{1}{L_1^2}\frac{1}{R_1^2}\frac{1}{S^2}\frac{1}{L_2^2}\frac{1}{R_2^2}\frac{1}{k_1^2}\frac{1}{k_2^2}\,,
\eeq
where the numerator is
\begin{equation}
\label{num2}
({\rm Num})_2=-2(x+\xi)^2\bigg\{\frac{-2\kb^2_1}{\beta_1}\bigg[1+\frac{2(x-\xi)}{\beta_1}\bigg]\bigg\}\bigg\{\frac{-2\kb^2_2}{\beta_2}\bigg[1+\frac{2(\beta_1+x-\xi)}{\beta_2}\bigg]\bigg\}2s \ ,
\end{equation}
and  the propagators are 
\begin{eqnarray}
L_1^2&=&\alpha_1(x+\xi)s \  \  \  , \  \  \  R_1^2=-\kb_1^2+\alpha_1(\beta_1+x-\xi)s\,, \nonumber \\ 
L_2^2&=&\alpha_2(x+\xi)s \  \  \  , \  \  \  R_2^2=-\kb_2^2+\alpha_2(\beta_1+\beta_2+x-\xi)s\,, \nonumber  \\
S^2& =& -\kb_2^2+(\beta_1+\beta_2+x-\xi)s \ .
\end{eqnarray}
We perform Cauchy integration over $\alpha_1$ and $\alpha_2$ taking the residue at the pole $k_1^2=k_2^2=0$. This leads to the result
\bea
Res_{\alpha_1,\alpha_2}&=&4s\frac{1}{x-\xi}
\bigg[1+\frac{2(x-\xi)}{\beta_1}\bigg]\bigg[1+\frac{2(\beta_1+x-\xi)}{\beta_2}\bigg]\nonumber \\
&&\hspace{3cm}\times\frac{1}{\kb_1^2}\frac{1}{(\beta_1+x-\xi)}\frac{1}{\kb_2^2}\frac{1}{\kb_2^2-(\beta_1+\beta_2+x-\xi)s}\,.
\eea
The integral $I_2$, Eq.~(\ref{def:I2}), is written as 
\bea
I_2&=&4s\frac{(2\pi i)^2}{x-\xi}\int_0^{\xi-x}d\beta_1\int_0^{\xi-x-\beta_1}d\beta_2\int_0^{\infty}d_2\kb_2\int_0^{\kb_2^2}d_2\kb_1 \nonumber \\
&& \hspace{4cm}\times \frac{1}{\beta_1+x-\xi}\frac{1}{\kb_1^2}\frac{1}{\kb_2^2}\frac{1}{\kb_2^2-(\beta_1+\beta_2+x-\xi)s} \ ,
\label{I2-two-loop-box}
\eea
with the limits of the $\beta$ integration determined according to the discussion in App.~\ref{Ap:beta-range}. Moreover, the expressions  $\bigg[1+\frac{2(x-\xi)}{\beta_1}\bigg]$ and $\bigg[1+\frac{2(\beta_1+x-\xi)}{\beta_2}\bigg]$ are both approximated to -1 according to the discussion after Eq.~(\ref{ELVg}).
Using the identity 
\begin{equation}
\label{identity2}
\int_0^{\kb_2^2}d_2\kb_1=\int_0^{\infty}d_2\kb_1-\int_{\kb_2^2}^{\infty}d_2\kb_1\,,
\end{equation}
which effectively shifts infrared divergences into ultraviolet divergences, one can use dimensional regularization for the first integral in $\kb_1$ which then vanishes since it is scaleless. Thus, only the second integral contributes and the two-loop integral is written as 
\bea
I_2&=&-4s\frac{(2\pi i)^2}{x-\xi}\int_0^{\xi-x}d\beta_1\int_0^{\xi-x-\beta_1}d\beta_2\int_0^{\infty}d_N\kb_2\int_{\kb_2^2}^{\infty}d_N\kb_1\nonumber \\
&& \hspace{4.5cm}\times\frac{1}{\beta_1+x-\xi}\frac{1}{\kb_1^2}\frac{1}{\kb_2^2}\frac{1}{\kb_2^2-(\beta_1+\beta_2+x-\xi)s} \ .
\eea
Using the identity (\ref{identity1}),
\beq
\frac{1}{\kb_2^2}\frac{1}{\kb_2^2-(\beta_1+\beta_2+x-\xi)s}=\bigg[\frac{1}{\kb_2^2}-\frac{1}{\kb_2^2-(\beta_1+\beta_2+x-\xi)s}\bigg]\frac{-1}{(\beta_1+\beta_2+x-\xi)s} \ , 
\eeq
the first term vanishes in dimensional regularization since there is no scale. Then the integral $I_2$ can be written as 
\bea
I_2&=&-4\frac{(2\pi i)^2}{x-\xi}\int_0^{\xi-x}d\beta_1\int_0^{\xi-x-\beta_1}d\beta_2\frac{1}{\beta_1+x-\xi}\frac{1}{\beta_1+\beta_2+x-\xi}\nonumber \\
&&\hspace{3cm}\times\int_0^{\infty}d_N\kb_2\int_{\kb_2^2}^{\infty}d_N\kb_1\frac{1}{\kb_1^2}\frac{1}{\kb_2^2-(\beta_1+\beta_2+x-\xi)s}\,.
\eea
Integrating over $\kb_1$ within dimensional regularization and taking only the finite term, we have
\bea
I^{\rm fin.}_2&=&4\frac{(2\pi i)^2}{x-\xi}\int_0^{\xi-x}d\beta_1\int_0^{\xi-x-\beta_1}d\beta_2\frac{1}{\beta_1+x-\xi}\frac{1}{\beta_1+\beta_2+x-\xi}\nonumber \\
&&\hspace{4cm}\times\int_0^{\infty}d_N\kb_2 \log{\kb_2^2}\frac{1}{\kb_2^2-(\beta_1+\beta_2+x-\xi)s} \ .
\eea
Integration over $\kb_2$ can be performed in the same way. At this point we note that the same result can be obtained without invoking explicitly dimensional regularization, but using the method that is described in the App.~\ref{Ap:integrals}. Thus, using Eq.~(\ref{result-integral}) we get
\bea
I^{\rm fin.}_2&=&-4\frac{(2\pi i)^2}{x-\xi}\int_0^{\xi-x}d\beta_1\int_0^{\xi-x-\beta_1}d\beta_2\frac{1}{\beta_1+x-\xi}\frac{1}{\beta_1+\beta_2+x-\xi}\nonumber \\
&&\hspace{7cm}\times\frac{1}{2!}\log^2(\xi-x-\beta_1-\beta_2) \ .
\eea
It is straightforward to integrate over $\beta_1$ and $\beta_2$. Using the matching condition with the exact one-loop result, the integral $I_2$ reads
\begin{equation}
I^{\rm fin.}_2=-4\frac{(2\pi i)^2}{x-\xi+i\epsilon}\frac{1}{4!}\log^4\bigg[\frac{\xi-x}{2\xi}-i\epsilon\bigg]
\,.
\end{equation}


\subsection{Detailed analysis of the suppressed diagrams at two-loop}
\label{SubSec:suppressed-2-loop}

In this section, we study in detail 
some of the two-loop diagrams in order to infer the minimal rules on which we will rely to then show that any diagram except the ladder-like one are suppressed. These rules will be enough to also justify in the next section that only ladder-like diagrams contribute at any order.

\begin{figure}[h!]
\centerline{
\psfrag{kg1}{\raisebox{.2cm}{\hspace{-1.2cm}$p_1 - 2 \xi \, p_2$}}
\psfrag{kg2}{\raisebox{.2cm}{\hspace{.3cm}$p_1$}}
\psfrag{delta}{$\,\,S$}
\psfrag{ki}{\!\hspace{-1.2cm}$(x+\xi)\, p_2$}
\psfrag{kf}{$(x-\xi)\, p_2$}
\psfrag{ku}{$k_2$}
\psfrag{kd}{$k_1$}
\psfrag{L1}{\!$L_1$}
\psfrag{L2}{\!$L_2$}
\psfrag{R1}{$R_1$}
\psfrag{R2}{$R_2$}
 \includegraphics[width=4.65cm]{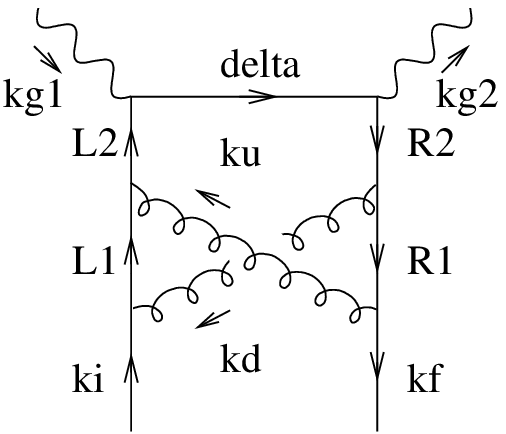} }
\caption{The two-loop subleading cross diagram.}
\label{Fig:2-loop-cross}
\end{figure}

Let us first consider the cross diagram illustrated in Fig.~\ref{Fig:2-loop-cross}. It reads
\beq
I=\int d^dk_1 d^dk_2 {\rm (Num)}\frac{1}{L_1^2}\frac{1}{L_2^2}\frac{1}{S^2}\frac{1}{R_1^2}\frac{1}{R_2^2}\frac{1}{k_1^2}\frac{1}{k_2^2} \ ,
\label{def-I-cross}
\eeq
with the numerator (Num) given by
\beq
{\rm (Num)}=\tr\bigg\{\ps_2\es_1\bigg(\frac{\beta_1}{2}+x-\xi\bigg)\ps_2\es_2\bigg(\frac{\beta_2}{2}+\beta+x-\xi\bigg)\ps_2\theta(x+\xi)\ps_2\es_1(x+\xi)\ps_2\es_2\bigg\}\,,
\label{num-cross-2-loop}
\eeq
where the denominators are
\bea
\label{deno-cross-2-loop}
L_1^2&=&-\kb_1^2+\alpha_1(\beta_1+x+\xi)s\,, \qquad L_2^2=-(\kb_1+\kb_2)^2+(\alpha_1+\alpha_2)(\beta_1+\beta_2+x+\xi)s \,, \nonumber \\
R_1^2&=&-\kb_2^2+\alpha_2(\beta_2+x-\xi)s \,,  \qquad R_2^2=-(\kb_1+\kb_2)^2+(\alpha_1+\alpha_2)(\beta_1+\beta_2+x-\xi)s \,, \nonumber \\
S^2&=&-(\kb_1+\kb_2)^2+(1+\alpha_1+\alpha_2)(\beta_1+\beta_2+x-\xi)s \,.
\eea
The ordering which leads to the dominant contribution is provided by a strong ordering both of transverse momenta and collinear momenta, to extract the maximal logarithmic contributions, as
\beq
|\kb_2| \gg |\kb_1| \quad {\rm and } \quad x \sim \xi \gg |\beta_1| \gg |\beta_2|
\label{ordering-momenta-cross}
\eq
(or $|\kb_2| \ll |\kb_1|$ and $x \sim \xi \gg |\beta_2| \gg |\beta_1|$). One can easily check by inspection that any other ordering leads to  less power  of logarithms. Using the ordering (\ref{ordering-momenta-cross}), the residue is
\bea
Res_{\alpha_1 , \alpha_2}&=&-4s\bigg[1+\frac{2(x-\xi)}{\beta_1}\bigg]\bigg[1+\frac{2(\beta_1+x-\xi)}{\beta_2}\bigg]\frac{\beta_2^{2}}{\beta_1(\beta_1+x-\xi)}\frac{1}{{\kb_2}^2(x-\xi)}\nonumber \\ && \hspace{3cm} \times\frac{1}{{\kb_2}^2(\beta_1+x-\xi)}\frac{1}{{\kb_2}^2-(\beta_1+\beta_2+x-\xi)s}\,,
\eea
and the integral to be computed is
\bea
\label{intcase1}
I&=&4s(2\pi i)^2\int_0^{\xi-x}d\beta_1 \int_0^{\xi-x-\beta_1}d\beta_2\int_0^{\infty}d_2\kb_2\int_0^{{\kb_2}^2}d_2\kb_1\nonumber\\ && \hspace{2cm}\times\frac{1}{x-\xi}\frac{1}{{\kb_2}^2(x-\xi)}\frac{1}{{\kb_2}^2}\frac{1}{{\kb_2}^2-(\beta_1+\beta_2+x-\xi)s}\,.
\eea
It is instructive to compare this expression with Eq.~(\ref{I2-two-loop-box}).
We explicitly see that integration over $\kb_1$ is different, since there is no $\frac{1}{\kb_1^2}$ appearing which is the source of one power of $\log(\xi-x)$. This is due to the fact that this cross diagram does not generate maximal collinear singularities. It thus shows that the net result can be neglected with respect to 
 the dominant contribution $\sim \frac{\log^4(\xi-x)}{(x-\xi)}$.

\begin{figure}[h!]

\psfrag{kg1}{}
\psfrag{kg2}{}
\psfrag{k2}{$\,k_2$}

\psfrag{k1}{\,$k_1$}

\psfrag{c1}{\raisebox{0cm}{\hspace{-.1cm}\rotatebox{90}{$\Large -$}}}
\psfrag{c2}{\raisebox{0cm}{\hspace{-.1cm}\rotatebox{90}{$\Large -$}}}
\psfrag{ki}{}
\psfrag{kf}{}

\psfrag{L1}{$\kb_1^2$}
\psfrag{L2}{$\kb_2^2$}

\psfrag{R1}{$\kb_1^2$}
\psfrag{R2}{$\kb_2^2$}

\psfrag{delta}{\hspace{-.2cm}$\kb_2^2+\Delta$}
\centerline{\includegraphics[width=5.3cm]{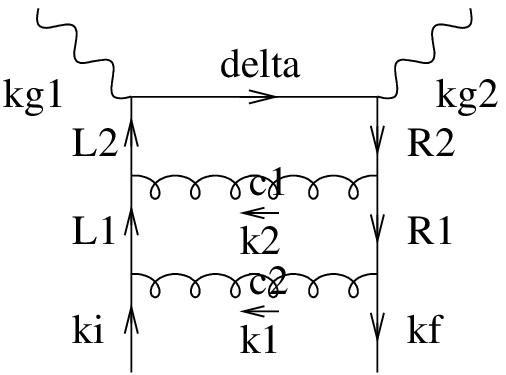} \hspace{.2cm}
\psfrag{kg1}{}
\psfrag{kg2}{}
\psfrag{k2}{$\,k_2$}
\psfrag{k1}{\,$k_1$}
\psfrag{c1}{\raisebox{0cm}{\hspace{-.1cm}\rotatebox{90}{$\Large -$}}}
\psfrag{c2}{\raisebox{0cm}{\hspace{-.1cm}\rotatebox{90}{$\Large -$}}}
\psfrag{ki}{}
\psfrag{kf}{}
\psfrag{L1}{$\kb_1^2$}
\psfrag{L2}{$\kb_1^2$}
\psfrag{R1}{$\kb_1^2$}
\psfrag{R2}{$\kb_1^2$}
\psfrag{delta}{\hspace{-.2cm}$\kb_1^2+\Delta$}
 \includegraphics[width=5.3cm]{two-loop-large-reverse-no-photon-arrows.eps}}
\caption{The two-loop ladder diagram which the two $\kb_i$ orderings. In both cases, the virtualities of the various propagators are indicated. Left: natural ordering $|\kb_2| \gg |\kb_1|\,,$ leading to the dominant contributions. Right: $|\kb_1| \gg |\kb_2|$ leading to a suppressed contribution.}
\label{Fig:2-loop-ladder-orderings}
\end{figure}

The same reasoning applies to the ladder-like diagram which we have discussed in Sec.~\ref{SubSec:ladder-2}. The Fig.~\ref{Fig:2-loop-ladder-orderings} shows 
 the virtualities of the various propagators in the two possible ordering in $\kb_i$, keeping the usual ordering in $\beta_i$, namely
$x \sim \xi\gg \vert \beta_1\vert  \sim \vert x-\xi \vert \gg \vert x-\xi +\beta_1\vert \sim \vert \beta_2 \vert$.
The left diagram, where the ordering  $|\kb_2| \gg |\kb_1|$ is assumed, exhibits a maximal number of collinear singularities while the right one, with the opposite ordering $|\kb_2| \ll |\kb_1|$, does not.
This justifies the $\kb_i$ ordering which was used in Sec.~\ref{SubSec:ladder-2}.

From the above study, we can now infer the second guiding rule to extract the leading contribution in powers of $\log(\xi-x)$, namely:
\\

\no
{\it (ii) Each loop should involve a maximal number of collinear singularities, which manifest themselves as maximal powers of $1/\kb_i^2$ for each $i$, after the $\alpha_i$ integration according to rule~(i).} 
\\

%
%

\begin{figure}[h!]
\centerline{\psfrag{kg1}{}
\psfrag{kg2}{}
\psfrag{k1}{\raisebox{.2cm}{\hspace{.1cm}$k_1$}}
\psfrag{k2}{\raisebox{.3cm}{$k_2$}}
\psfrag{k3}{$\hspace{-.25cm}\kb_2^2+\Delta$}
\psfrag{k4}{\hspace{-.3cm}$\kb_2^2+\Delta'$}
\psfrag{ki}{}
\psfrag{kf}{}
\psfrag{R1}{$\kb_2^2$}
\psfrag{R2}{$\kb_1^2$}
\psfrag{L1}{$\kb_2^2$}
\psfrag{c1}{\raisebox{.5cm}{\hspace{-.05cm}\rotatebox{90}{$\Large -$}}}
\psfrag{c2}{\raisebox{.17cm}{\hspace{.2cm}\rotatebox{35}{$\Large -$}}}
 \scalebox{.9}{\includegraphics[width=7cm]{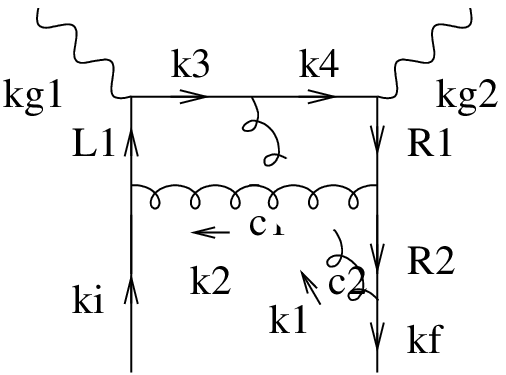}}
 \qquad
\psfrag{k4}{\hspace{-.3cm}$\kb_1^2+\Delta'$}
\psfrag{R1}{$\kb_1^2$} \scalebox{.9}{\includegraphics[width=7cm]{two-loop-cross-rightFULL-reverse-no-arrows-photons-cut-gluons.eps}}}
\caption{A two-loop subleading diagram with a gluon attached to the $s-$channel fermionic line. The virtualities of fermionic lines are indicated. Left: ordering $\kb_2^2 \gg \kb_1^2.$
Right: ordering $\kb_2^2 \ll \kb_1^2.$
}
\label{Fig:2-loop-cross-s-channel-right-orderings}
\end{figure}

We now consider the diagram of Fig.~\ref{Fig:2-loop-cross-s-channel-right-orderings},
which involves the coupling of a gluon to the $s-$channel fermionic line.
Closing the $\alpha_i$ contours on the gluonic poles, 
the fermionic propagators get virtualities whose order of magnitude are indicated on  Fig.~\ref{Fig:2-loop-cross-s-channel-right-orderings}.
Two limits are of interest in order to obtain the maximal powers of $\log (\xi -x)$. The first one is the limit $\kb_2^2 \gg \kb_1^2.$ In that case, the number of collinear singularities originating from $k_1$ is too low, since there is a single propagator of virtuality $\kb_1^2$ (compensated by a similar $\kb_1^2$ in the numerator), and this contribution is subleading.
The second one is the limit $\kb_2^2 \ll \kb_1^2.$ In this case, the fact that the upper left fermionic propagator has a virtuality $\kb_2^2 + \Delta$ where $\Delta=-(x-\xi+\beta_2)s$ lowers the level of singularity, again leading to a suppressed contribution.

From this study, we can now infer the third guiding rule to extract the leading contribution in powers of $\log(\xi-x)$, namely:
\\

\no
{\it (iii)  Any coupling of a gluon to the $s-$channel fermionic line leads to a suppressed contribution.}
\\

\begin{figure}[h!]
\centerline{
\psfrag{delta}{$\!\!\!\kb_1^2+\Delta$}
\psfrag{kg1}{}
\psfrag{kg2}{}
\psfrag{ki}{}
\psfrag{kf}{}
\psfrag{R1}{$\kb_1^2$}
\psfrag{R2}{$\kb_2^2$}
\psfrag{R3}{$\kb_1^2$}
\psfrag{L1}{$\kb_1^2$}
\psfrag{k1}{\hspace{-.25cm}$k_1$}
\psfrag{k}{\raisebox{-.015cm}{$\hspace{-.7cm}{\Large -} \hspace{.3cm} k_2$}}
\psfrag{c1}{\raisebox{.6cm}{\hspace{-.02cm}\rotatebox{90}{$\Large -$}}}
\includegraphics[width=4.5cm]{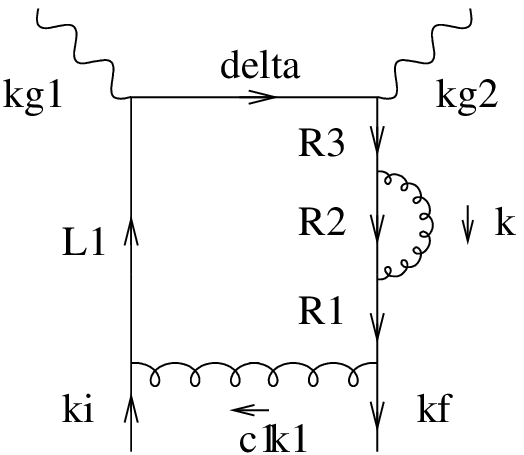}
\qquad
\psfrag{delta}{$\!\!\!\kb_1^2+\Delta$}
\psfrag{kg1}{}
\psfrag{kg2}{}
\psfrag{ki}{}
\psfrag{kf}{}
\psfrag{R1}{$\kb_1^2$}
\psfrag{R2}{$\kb_1^2$}
\psfrag{R3}{$\kb_1^2$}
\psfrag{L1}{$\kb_1^2$}
\psfrag{k1}{\hspace{-.25cm}$k_1$}
\psfrag{k}{\raisebox{-.015cm}{$\hspace{-.7cm}{\Large -} \hspace{.3cm} k_2$}}
\psfrag{c1}{\raisebox{.6cm}{\hspace{-.02cm}\rotatebox{90}{$\Large -$}}}
\includegraphics[width=4.5cm]{two-loop-right-reverse-no-arrows-photons.eps}
}
\caption{A subleading two-loop diagram of abelian type. The virtualities of the propagators are indicated when closing on the gluonic poles. Left: ordering $\kb_2^2 \gg \kb_1^2.$
Right: ordering $\kb_2^2 \ll \kb_1^2.$}
\label{Fig:2-loop-abelian}
\end{figure}

The last rule is obtained through the study of a diagram involving a fermion self-energy, of the type shown in Fig.~\ref{Fig:2-loop-abelian}.
The key point here is to realise that the virtuality of the $s-$channel fermion is $\kb_1^2+\Delta$, where $\Delta=-(x-\xi+\beta_1)s\,.$ The fact that $\Delta$ does not involve $\beta_2$ reduces the power of $\log (\xi-x)$ when integrating over 
$\beta_2$.

From this study, we can now infer the fourth guiding rule to extract the leading contribution in powers of $\log(\xi-x)$, namely:
\\

\no
{\it (iv) The diagram should be sufficiently non-local in order that the $s-$channel fermionic line involves the whole $p_2$ flux.}
\\

\begin{figure}[h!]
\psfrag{kg1}{}
\psfrag{kg2}{}
\psfrag{k1}{}
\psfrag{k2}{}
\psfrag{ku}{}
\psfrag{kd}{}
\psfrag{ki}{}
\psfrag{kf}{}
\psfrag{delta}{}
\psfrag{k}{}
\psfrag{R1}{}
\psfrag{R2}{}
\psfrag{L1}{}
\psfrag{L2}{}
\centerline{\ \includegraphics[width=4cm]{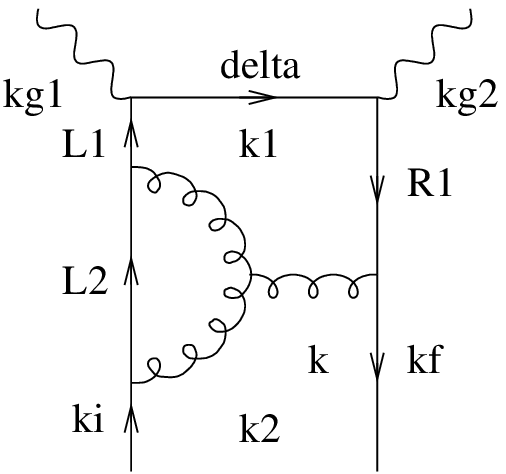} \qquad  \includegraphics[width=4cm]{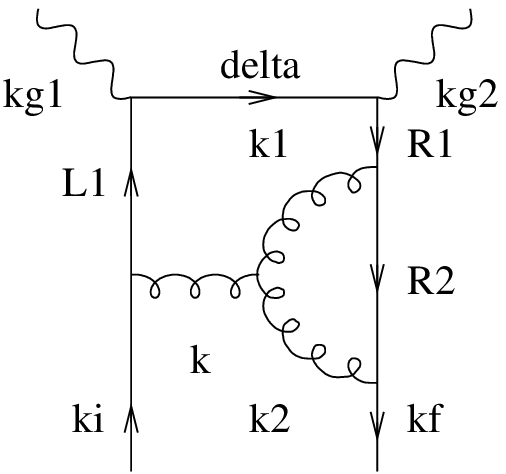}
\qquad
\includegraphics[width=4.4cm]{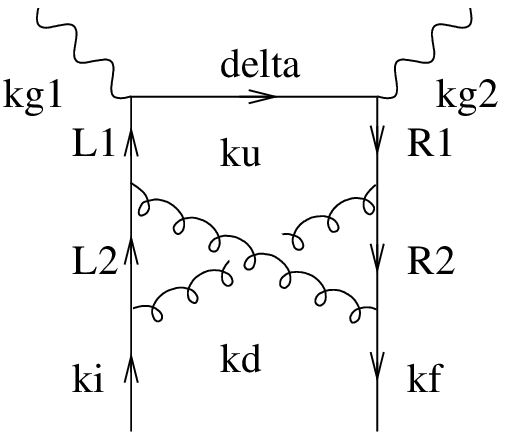}
}
\caption{Subleading two-loop diagrams violating rule (ii).}
\label{Fig:2-loop-rule-2}
\end{figure}

\begin{figure}[h!]
\psfrag{kg1}{}
\psfrag{kg2}{}
\psfrag{k1}{}
\psfrag{k2}{}
\psfrag{k3}{}
\psfrag{k4}{}
\psfrag{ku}{}
\psfrag{kd}{}
\psfrag{ki}{}
\psfrag{kf}{}
\psfrag{delta}{}
\psfrag{k}{}
\psfrag{R1}{}
\psfrag{R2}{}
\psfrag{L1}{}
\psfrag{L2}{}
\centerline{\raisebox{0cm}{\includegraphics[width=4cm]{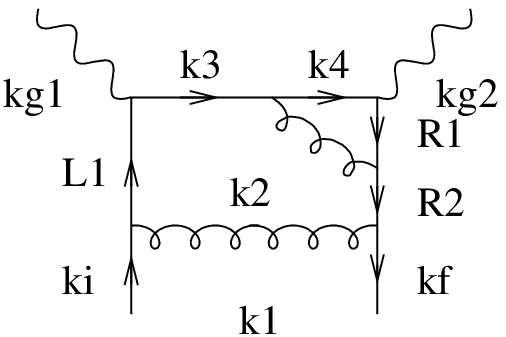}} \quad
\includegraphics[width=4cm]{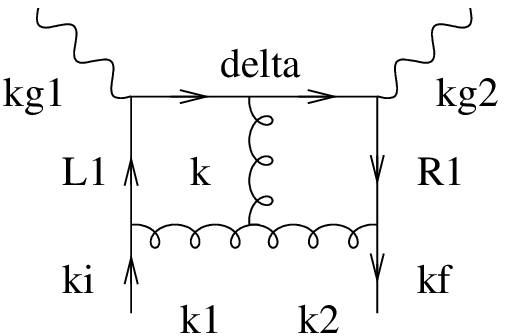} \quad
\includegraphics[width=4cm]{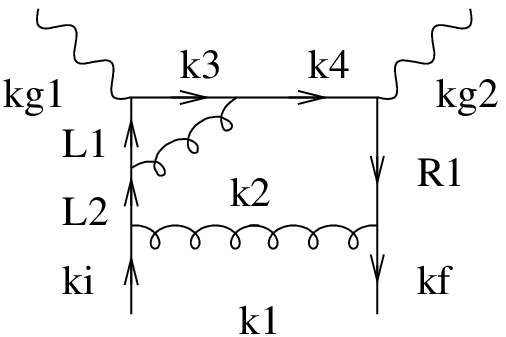}}
\vspace{.2cm}

\centerline{\includegraphics[width=3.5cm]{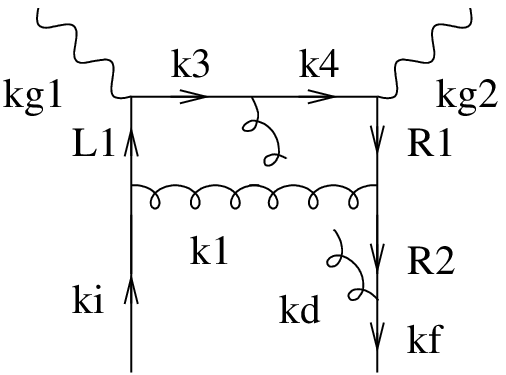}
\quad \qquad
\includegraphics[width=3.5cm]{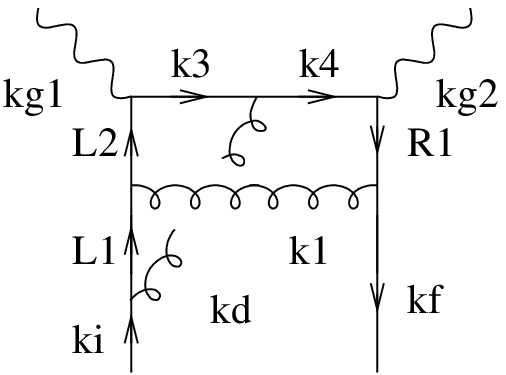}
}
%
%
\caption{Subleading two-loop diagrams violating rule (iii).}
\label{Fig:2-loop-rule-3}
\end{figure}

\begin{figure}[h!]
\centerline{
\psfrag{kg1}{}
\psfrag{kg2}{}
\psfrag{k1}{}
\psfrag{k2}{}
\psfrag{k3}{}
\psfrag{k4}{}
\psfrag{ku}{}
\psfrag{kd}{}
\psfrag{ki}{}
\psfrag{kf}{}
\psfrag{delta}{}
\psfrag{k}{}
\psfrag{R1}{}
\psfrag{R2}{}
\psfrag{R3}{}
\psfrag{L1}{}
\psfrag{L2}{}
\psfrag{L3}{}
\includegraphics[width=4cm]{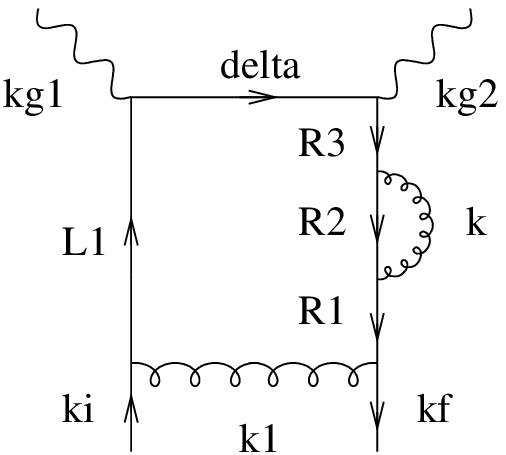} \quad \quad \quad
\includegraphics[width=4cm]{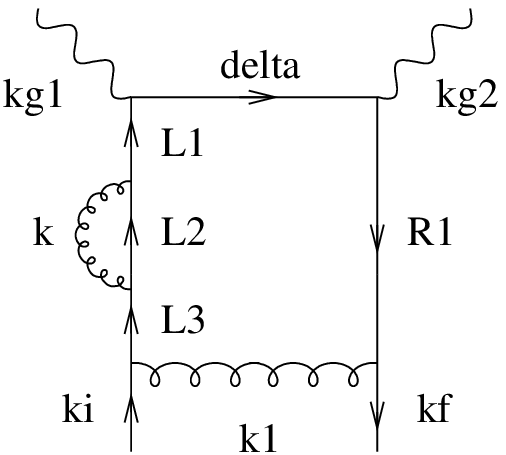}}
\vspace{.2cm}

\centerline{
\psfrag{kg1}{}
\psfrag{kg2}{}
\psfrag{k1}{}
\psfrag{k2}{}
\psfrag{k3}{}
\psfrag{k4}{}
\psfrag{ku}{}
\psfrag{kd}{}
\psfrag{ki}{}
\psfrag{kf}{}
\psfrag{delta}{}
\psfrag{k}{}
\psfrag{R1}{}
\psfrag{R2}{}
\psfrag{R3}{}
\psfrag{L1}{}
\psfrag{L2}{}
\psfrag{L3}{}
\includegraphics[width=4cm]{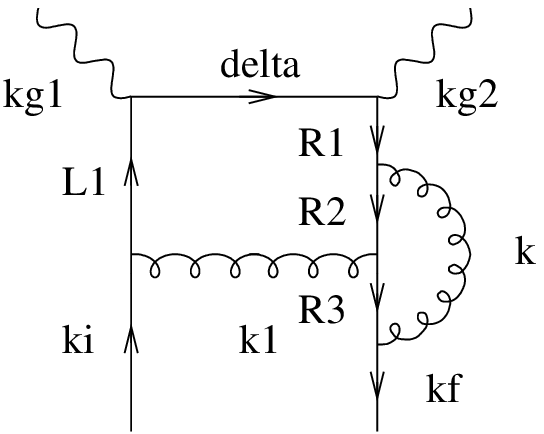} \quad \quad 
\includegraphics[width=4cm]{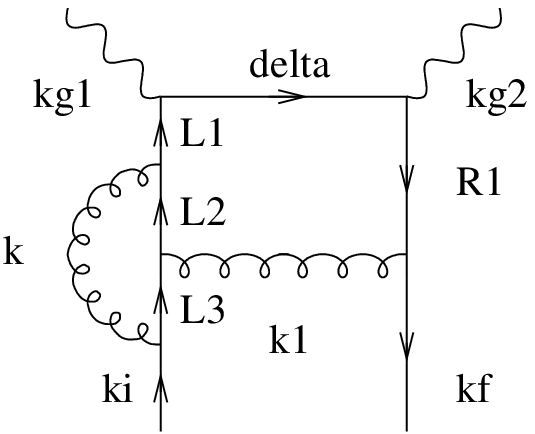}}
\caption{Subleading two-loop diagrams violating rule (iv).}
\label{Fig:2-loop-rule-4}
\end{figure}

These four rules are sufficient to show that any non ladder-like diagram is suppressed, as we show now. The 3 diagrams of Fig.~\ref{Fig:2-loop-rule-2}
are suppressed after applying rule (ii). The 5 diagrams of Fig.~\ref{Fig:2-loop-rule-3}
are suppressed after applying rule (iii). And the 4 diagrams of Fig.~\ref{Fig:2-loop-rule-4}
are suppressed after applying rule (iv). This last rule also excludes diagrams with 
virtual corrections on the gluon propagator.

%

\section{All-loop analysis}
\label{Sec:n-order}

\subsection{Beyond the two-loop order}

Based on the four rules formulated previously, it is now possible to justify that the contributions to the maximal powers of $\frac{\log^{2n}(\xi-x)}{x-\xi}$ only arise from the ladder-like diagram, at any order $\alpha_s^n$ . For that, we are using a recursive argument. 
At two loop, we have seen that the diagrams with 3-gluon coupling are subdominant, since the powers of $\kb_i$ are not maximal in that case. At three loop, the last missing building block, namely the 4-gluon vertex, appears. Since it is a contraction of two three-loop diagrams (which are already excluded) with one less propagator, this kind of vertex is also subleading. 

Thus, starting from the ladder-like diagram at order $n-1$, let us dress it having in mind that we are looking for the maximal power of 
$\log(\xi-x)\,,$ which will look ultimately like $\frac{1}{x-\xi}\log^{2 n}(\xi-x)$.
First, we are only allowed to consider abelian-like diagrams. Starting from a gluon which is attached somewhere on the right fermionic line, this line should end 
up on the left fermionic line: ending on the right would be too local (rule (iv)), and ending on the $s-$channel fermionic line would violate rule (iii). Finally, a crossing of any gluon line is not permitted since the rule (ii) would not be satisfied. Thus, we end up with the ladder-like diagram of order $n.$

\subsection{Computation of the $n$-loop ladder diagram}
\label{SubSec:n-loop-computation}

The computation of the $n$-loop ladder diagram (see Fig.~\ref{Fig:n-loop}) is performed in full analogy with the one- and two- loop diagrams discussed in Sections \ref{Sec:1-order-ladder} and \ref{SubSec:ladder-2}. We again assume that all rung gluons in the ladder diagram are on the mass shell, and that there is a strong ordering in their transverse momenta and Sudakov variables $\alpha_i$, $\beta_i$, i.e.
\bea
&&\hspace{-.2cm}|\kb_n| \gg |\kb_{n-1}| \gg \cdots \gg  |\kb_1|\  \  \  \  \  ,  \  \  \  \  \ 1 \gg |\alpha_n| \gg |\alpha_{n-1}|\gg \cdots \gg|\alpha_1|\,, \\
&&\hspace{-.25cm} x \sim \xi\gg \vert \beta_1\vert  \sim \vert x-\xi \vert \gg \vert x-\xi +\beta_1\vert \sim \vert \beta_2 \vert \gg  \!\cdots  \!\!\gg   \vert x-\xi +\beta_1 +\beta_2 -\cdots+ \beta_{n-1} \vert  \sim  \vert \beta_n  \vert  . \nonumber
\eea
The above assumptions permit us to have eikonal coupling on the left fermionic line and on the right fermionic line the coupling goes beyond the eikonal coupling  taking into account some recoil effects. Thus, the integral $I_n$ for the n-loop ladder diagram is written as 
\beq
I_{n}=\bigg(\frac{s}{2}\bigg)^n\int d\alpha_1\, d\beta_1\, d_2\kb_1\cdots\int d\alpha_n\, d\beta_n\, d_2\kb_n~({\rm Num})_n\frac{1}{L^2_1}\cdots\frac{1}{L^2_n} \frac{1}{S^2}\frac{1}{R^2_1}\cdots\frac{1}{R^2_n}\frac{1}{k_1^2}\cdots\frac{1}{k_n^2}\,,
\eeq
where the numerator $({\rm Num})_n$ takes the form (compare with Eqs.~(\ref{numfinal}) and (\ref{num2}))
\bea
&&({\rm Num})_n=-2(x+\xi)^n\bigg\{\frac{-2\kb_1^2}{\beta_1}\bigg[1+\frac{2(x-\xi)}{\beta_1}\bigg]\bigg\}\bigg\{\frac{-2\kb_2^2}{\beta_2}\bigg[1+\frac{2(\beta_1+x-\xi)}{\beta_2}\bigg]\bigg\} \nonumber \\
&&\hspace{5cm} \cdots \bigg\{\frac{-2\kb_n^2}{\beta_n}\bigg[1+\frac{2(\beta_{n-1}+\cdots +\beta_1x-\xi)}{\beta_n}\bigg]\bigg\}2s\,,
\eea
with the propagators 
\begin{eqnarray}
L_1^2&=&\alpha_1(x+\xi)s \    ,\  \  \  \  \  R_1^2=-\kb_1^2+\alpha_1(\beta_1+x-\xi)s\,, \nonumber \\ 
L_2^2&=&\alpha_2(x+\xi)s \   , \  \  \  \  \  R_2^2=-\kb_2^2+\alpha_2(\beta_1+\beta_2+x-\xi)s \,,\nonumber  \\
&\vdots& \nonumber \\
L_n^2&=&\alpha_n(x+\xi)s \  , \  \  \  \  \  R_n^2=-\kb_n^2+\alpha_n(\beta_1+\cdots +\beta_n+x-\xi)s\,,\nonumber \\ 
S^2& =&-\kb_n^2+(\beta_1+\cdots+\beta_n+x-\xi)s\,.
\end{eqnarray}
Calculating the residue in $\alpha_i$, we get
\begin{eqnarray}
&&Res_{\alpha_1,\cdots \alpha_n}=4s\frac{1}{x-\xi}\frac{1}{(\beta_1+x-\xi)}\cdots\frac{1}{\beta_1+\cdots+\beta_{n-1}+x-\xi}\nonumber \\
&& \hspace{6cm}\times\frac{1}{\kb_1^2}\cdots\frac{1}{\kb_n^2}\frac{1}{\kb_n^2-(\beta_1+\cdots+\beta_n+x-\xi)s} \ ,
\end{eqnarray}
where each expression $\bigg[1+\frac{2(x-\xi)}{\beta_1}\bigg]$, $\bigg[1+\frac{2(\beta_1+x-\xi)}{\beta_2}\bigg]$, $\cdots$ and $\bigg[1+\frac{2(\beta_{n-1}+\cdots +\beta_1x-\xi)}{\beta_n}\bigg]$ is approximated to -1 according to the discussion after Eq.~(\ref{ELVg}).
Then the integral $I_n$ reads
\begin{eqnarray}
I_n&=&4s(-1)^n\frac{(2\pi i)^n}{x-\xi}\int_0^{\xi-x}\!           \!d\beta_1\cdots\!\int_0^{\xi-x-\beta_1-\cdots-\beta_{n-1}}\hspace{-.6cm}d\beta_n\frac{1}{\beta_1+x-\xi}\cdots\frac{1}{\beta_1+\cdots+\beta_{n-1}+x-\xi}\nonumber\\
&&\hspace{2cm}\times \int_0^{\infty}d_2\kb_n\cdots\int_0^{\kb_2^2}d_2\kb_1\frac{1}{\kb_1^2}\cdots\frac{1}{\kb_n^2}\frac{1}{\kb_n^2-(\beta_1+\cdots+\beta_n+x-\xi)s} \ ,
\end{eqnarray}
in which the limits of the $\beta_i$ integrations are determined according to the discussion in App.~\ref{Ap:beta-range}.
Now we will use dimensional regularization to integrate over momenta $\kb_i$. We use the identity (\ref{identity2}) for the integral over $\kb_1$. The first integral on the right hand side of the identity~(\ref{identity2}) vanishes. Using the same argument for $n-1$ momenta and using the identity (\ref{identity1}) for the product of the terms with momentum $\kb_n$, the integral $I_n$ can be written as
\begin{eqnarray}
I_n&=&-4\frac{(2\pi i)^n}{x-\xi}\int_0^{\xi-x}d\beta_1\cdots\int_0^{\xi-x-\beta_1-\cdots-\beta_{n-1}}d\beta_n\frac{1}{\beta_1+x-\xi}\cdots\frac{1}{\beta_1+\cdots+\beta_n+x-\xi}\nonumber\\
&&\hspace{1cm}\times \int_0^{\infty}d_N\kb_n\cdots\int_{\kb_2^2}^{\infty}d_N\kb_1\frac{1}{\kb_1^2}\cdots\frac{1}{\kb_{n-1}^2}\frac{1}{\kb_n^2-(\beta_1+\cdots+\beta_n+x-\xi)s}\,.
\end{eqnarray}
Integrating over $\kb_1\,, \cdots,\, \kb_{n-1}$ and only retaining  the finite contributions from each integral, we  have
\bea
I^{\rm fin.}_n&=&-4\frac{(2\pi i)^n}{x-\xi}\int_0^{\xi-x}d\beta_1\cdots\int_0^{\xi-x-\beta_1-\cdots-\beta_{n-1}}d\beta_n\frac{1}{\beta_1+x-\xi}\cdots\frac{1}{\beta_1+\cdots+\beta_n+x-\xi}\nonumber\\
&&\hspace{2cm}\times \int_0^{\infty}d_N\kb_n(-1)^{n-1}\frac{1}{(n-1)!}\frac{\log^{n-1}\kb^2_n}{\kb_n^2-(\beta_1+\cdots+\beta_n+x-\xi)s}\,.
\eea
The result of the last integral over $\kb_n$ is given by Eq.~(\ref{result-integral}) in App.~\ref{Ap:integrals}. In this way we obtain
\bea
I^{\rm fin.}_n&=&-4\frac{(2\pi i)^n}{x-\xi}\int_0^{\xi-x}d\beta_1\cdots\int_0^{\xi-x-\beta_1-\cdots-\beta_{n-1}}d\beta_n\frac{1}{\beta_1+x-\xi}\cdots\frac{1}{\beta_1+\cdots+\beta_n+x-\xi}\nonumber\\
&&\hspace{6cm}\times(-1)^n\frac{1}{n!}\log^n(\xi-x-\beta_1-\cdots-\beta_n)\,.
\eea
Keeping in mind the remarks about the dominant region of $\beta_i$ integrations after Eq.~(\ref{regI1}) and using the matching condition with the exact one-loop result,  the $\beta_i$ integrations lead to 
\begin{equation}
\label{in-final}
I^{\rm fin.}_n=-4\frac{(2\pi i)^n}{x-\xi+i\epsilon}\frac{1}{(2n)!}\log^{2n}\bigg[\frac{\xi-x}{2\xi}-i\epsilon\bigg]\,.
\end{equation}


\subsection{The resummed formula} 
\label{SubSec:resummed}

The Eqs.~(\ref{def-Kn}) and (\ref{in-final}) permits us to perform the resummation of the ladder diagrams and we obtain
\beq
\label{Sumn-final}
\left(\sum_{n=0}^\infty K_n \right) \, - \, (x \rightarrow -x)  =
\frac{e_q^2}{x-\xi + i \epsilon} \cosh
\left[ D \log \left( \frac{\xi -x}{2 \xi } - i \epsilon \right) \right] - \, (x \rightarrow -x) \,.
\eeq
The resummed to all orders formula Eq.~(\ref{Sumn-final}) can now be included into
the NLO coefficient function Eq.~(\ref{tq}). The inclusion procedure is not unique and it is natural to propose two  choices. The first case corresponds to modifying only the Born term  and $\log^2$ part of Eq.~(\ref{C1}) and keeping the rest of the terms unchanged. This corresponds to the following expression
\begin{eqnarray}
\label{Res1}
&&\hspace{-0.3cm}(T^q)^{\rm res1}=\bigg(\frac{e_q^2}{x-\xi+i\epsilon}\bigg\{
\cosh\bigg[D\log\bigg(\frac{\xi-x}{2\xi}-i\epsilon\bigg)\bigg]
-\frac{D^2}{2}\bigg[9+3\frac{\xi-x}{x+\xi}\log\bigg(\frac{\xi-x}{2\xi}-i\epsilon \bigg)\bigg]\bigg\}\nonumber \\
&&\hspace{8cm}+\,C^q_{coll}\log \frac{Q^2}{\mu_F^2}\bigg)-(x\rightarrow -x)\,.
\end{eqnarray}
In the second case the resummation effects are accounted for in a multiplicative way for $C_0^q$ and $C_1^q$, i.e. the resummed formula takes the following form
\begin{eqnarray}
\label{Res2}&&\hspace{-0.3cm}(T^q)^{\rm res2}=\bigg(\frac{e_q^2}{x-\xi+i\epsilon}\cosh\bigg[D\log\bigg(\frac{\xi-x}{2\xi}\!-\!i\epsilon\bigg)\bigg]\bigg[
1-\frac{D^2}{2}\bigg\{9+3\frac{\xi-x}{x+\xi}\log\bigg(\frac{\xi-x}{2\xi}-i\epsilon\bigg)\!\bigg\}\bigg]\nonumber \\
&&\hspace{8cm}+C^q_{coll}\log \frac{Q^2}{\mu_F^2}\bigg)-(x\rightarrow -x) \,,
\end{eqnarray}
where $D= \sqrt{\frac{\alpha_s C_F}{2\pi}}$.

These resummed formulas differ through logarithmic contributions which are beyond the precision of our study.






%

\section{Conclusions}
\label{Sec:conclusion}

The resummation of soft-collinear gluon radiation effects allowed us to get a close all-order formula that modifies
significantly the coefficient function in the specific region $x$ near $\pm \xi$. The measurement of the phenomenological
impact of this procedure on the data analysis needs further analysis with the implementation of modeled generalized
parton distributions and the discussion of specific observables. Let us just remind the reader that the region $x=\pm
\xi$ is crucial in the determination of  beam spin asymmetries. 

We did not study the case of gluon GPD contributions to DVCS, which, although they are absent at Born order,  are
expected \cite{Pire:2008ea} to become important in the small $\xi$ regime which will be accessible at high energies \cite{Boer:2011fh, Bruening:2012cc}.

Deeply virtual Compton scattering is but one of the exclusive processes giving access to GPDs. Our analysis could and
should be applied to other processes too. The case of timelike Compton scattering is special since, thanks to the
analyticity properties in $Q^2$, it has been shown \cite{Muller:2012yq}  that a simple relation was relating its NLO
correction to the one for DVCS. The case of exclusive meson production is also very interesting, both theoretically and
experimentally. The NLO analysis \cite{Belitsky:2001nq, Ivanov:2004zv} of the corresponding coefficient functions exhibit also a
$\log^2[(\xi-x)/2\xi] $ behavior, both in the quark and in the gluon channels. It will be most interesting to see if our
semi-eikonal analysis allows to resum these logarithms too. The quality of the present and near future data for vector
mesons (and in particular $\rho$) electroproduction demands this analysis to be vigorously pursued. 

We did not study the effects of the running of $\alpha_s$. 
Also, a formulation of resummation in our exclusive case in terms of (conformal) moments is not yet available. This would generalize analogous resummation of inclusive DIS cross-section which were performed in terms of Mellin moments. We leave studies of  these issues for a future work.

\section*{Acknowledgements}

We thank S.~Forte, A.~H.~Mueller, G.~Sterman, O.~V.~Teryaev, J.~Wagner,    for useful discussions and correspondence.
This work is partly supported by  the Polish Grant NCN No DEC-2011/01/D/ST2/02069, the French-Polish collaboration agreement Polonium, the P2IO consortium
 and the Joint
Research Activity "Study of Strongly Interacting Matter"
(acronym HadronPhysics3, Grant Agreement n.283286) under the
Seventh Framework Programme of the European Community.



\begin{appendix}
\section{Appendices}

\subsection{Extracting the $\beta_i$-ranges from the positions of the poles in $\alpha_i$}
\label{Ap:beta-range}

In this appendix we show how the $\beta_i$ integration range is obtained from the study of the position of $\alpha_i$ poles. 

Let us start with the $\alpha_1$ pole. The tree denominators $k_1^2$, $L_1$ and $R_1$ possesses simple zeros in $\alpha_1$. Due to the $i \epsilon$ prescription, their position in the complex-$\alpha_1$ plane is governed by their imaginary part. For these three denominators, we respectively get
\beqa
\label{position-pole-alpha1-1}
k_1^2&:& \ -\frac{i \epsilon}{\beta_1}\,, \\
\label{position-pole-alpha1-L1}
L_1&:& \ -\frac{i \epsilon}{\beta_1+x+\xi}\,, \\
\label{position-pole-alpha1-R1}
R_1&:& \ -\frac{i \epsilon}{\beta_1+x-\xi} \,.
\eqa
\def\up{\raisebox{.4cm}{\scalebox{1}[1.5]{\rotatebox{-90}{$\curvearrowleft$}}}}
\def\do{\raisebox{.4cm}{\scalebox{1}[1.5]{\rotatebox{-90}{$\curvearrowright$}}}}
Without loss of generality we can choose to close the $\alpha_1$ contour in order to always avoid the pole due to $R_1$. Since the dominant contribution comes from closing on the pole of $1/k_1^2$, this gives the 4 following possibilities out of 6
\bea
\label{pole-alpha1-1}
\frac{k_1 \ \phantom{L_1} \ \phantom{R_1} \ \up 
\hspace{-.3cm} 
}{\phantom{k_1} \ L_1 \ R_1} \\
\nonumber\\
\label{pole-alpha1-2}
\frac{\phantom{k_1} \ L_1 \ R_1}{k_1 \ \phantom{L_1} \ \phantom{R_1} \ \do
\hspace{-.3cm} }\\
\label{pole-alpha1-3}
\frac{\phantom{k_1} \ L_1 \ \phantom{R_1}}{k_1 \ \phantom{L_1} \ R_1}
\\
\nonumber\\
\label{pole-alpha1-4}
\frac{k_1 \ \phantom{L_1} \ R_1}{\phantom{k_1} \ L_1 \ \phantom{R_1}}\\
\nonumber\\
\label{pole-alpha1-5}
\frac{\phantom{k_1} \ \phantom{L_1} \ R_1}{k_1 \ L_1 \ \phantom{R_1}\ \do
\hspace{-.3cm}}\\
\nonumber\\
\label{pole-alpha1-6}
\frac{k_1 \ L_1 \ \phantom{R_1}\ \up
\hspace{-.3cm}}{\phantom{k_1} \ \phantom{L_1} \ R_1}%
\eqa
Combining (\ref{pole-alpha1-1}) and (\ref{pole-alpha1-6}), the restriction due to the position of the $1/L_1^2$ pole in complex plane disappear. Similarly, one can combine (\ref{pole-alpha1-2}) and (\ref{pole-alpha1-5}).
Thus one ends up with 
\bea
\label{pole-alpha1-a}
\frac{k_1 \  \phantom{R_1} \ \up 
\hspace{-.3cm} 
}{\phantom{k_1} \ R_1} && \qquad \Longrightarrow \ \phantom{-} \theta(-\beta_1) \, \theta(\beta_1+x-\xi) \\
\nonumber\\
\label{pole-alpha1-b}
\frac{\phantom{k_1} \  R_1}{k_1 \ \phantom{R_1} \ \do
\hspace{-.3cm}}&& \qquad \Longrightarrow \ -\theta(\beta_1) \, \theta(-\beta_1-x+\xi) \,,
\eqa
where we have used Eqs.~(\ref{position-pole-alpha1-1}, \ref{position-pole-alpha1-R1}).
Finally, the $\beta_1$ integral symbolicaly thus reads 
\beq
\label{results-beta1-spectrum}
-2 \pi i \int^{\xi-x}_0 d \beta_1 \, {\rm Res}_{\alpha_1}\,.
\eq
The same analysis can be applied for the $\alpha_2$ poles. 
Their position in the complex $\alpha_2$-plane is governed by
\beqa
\label{position-pole-alpha2-2}
k_2^2&:& \ -\frac{i \epsilon}{\beta_2} \,,\\
\label{position-pole-alpha2-L2}
L_2&:& \ -\frac{i \epsilon}{\beta_1+\beta_2+x+\xi}\,, \\
\label{position-pole-alpha2-R2}
R_2&:& \ -\frac{i \epsilon}{\beta_1+\beta_2+x-\xi} \,.
\eqa
and following the same line of thought we 
 choose to close the $\alpha_2$ contour in order to always avoid the pole due to $R_2$. This gives the 4 following possibilities out of 6
\bea
\label{pole-alpha2-1}
\frac{k_2 \ \phantom{L_2} \ \phantom{R_2} \ \up 
\hspace{-.3cm} 
}{\phantom{k_2} \ L_2 \ R_2} \\
\nonumber\\
\label{pole-alpha2-2}
\frac{\phantom{k_2} \ L_2 \ R_2}{k_2 \ \phantom{L_2} \ \phantom{R_2} \ \do
\hspace{-.3cm} }\\
\label{pole-alpha2-3}
\frac{\phantom{k_2} \ L_2 \ \phantom{R_2}}{k_2 \ \phantom{L_2} \ R_2}
\\
\nonumber\\
\label{pole-alpha2-4}
\frac{k_2 \ \phantom{L_2} \ R_2}{\phantom{k_2} \ L_2 \ \phantom{R_2}}\\
\nonumber\\
\label{pole-alpha2-5}
\frac{\phantom{k_2} \ \phantom{L_2} \ R_2}{k_2 \ L_2 \ \phantom{R_2}\ \do
\hspace{-.3cm}}\\
\nonumber\\
\label{pole-alpha2-6}
\frac{k_2 \ L_2 \ \phantom{R_2}\ \up
\hspace{-.3cm}}{\phantom{k_2} \ \phantom{L_2} \ R_2}%
\eqa
Combining (\ref{pole-alpha2-1}) and (\ref{pole-alpha2-6}), the restriction due to the position of the $1/L_2^2$ pole  in complex plane disappear. Similarly, one can combine (\ref{pole-alpha2-2}) and (\ref{pole-alpha2-5}).
One thus ends up with 
\bea
\label{pole-alpha2-a}
\frac{k_2 \  \phantom{R_2} \ \up 
\hspace{-.3cm} 
}{\phantom{k_2} \ R_2} && \qquad \Longrightarrow \ \phantom{-} \theta(-\beta_2) \, \theta(\beta_1+\beta_2+x-\xi) \\
\nonumber\\
\label{pole-alpha2-b}
\frac{\phantom{k_2} \  R_2}{k_1 \ \phantom{R_2} \ \do
\hspace{-.3cm}}&& \qquad \Longrightarrow \ -\theta(\beta_2) \, \theta(-\beta_1-\beta_2-x+\xi) \,,
\eqa
where we have used Eqs.~(\ref{position-pole-alpha2-2}, \ref{position-pole-alpha2-R2}).
Finally, the $\beta_2$ integral symbolicaly thus reads 
\beq
\label{results-beta2-spectrum}
-2 \pi i \int^{\xi-x-\beta_1}_0 d \beta_2 \, {\rm Res}_{\alpha_2}\,.
\eq
We apply the same analysis up to the $\alpha_{n-1}$ poles, which 
position in the complex $\alpha_{n-1}$-plane is governed by
\beqa
\label{position-pole-alpha_n-1-1}
k_{n-1}^2&:& \ -\frac{i \epsilon}{\beta_{n-1}}\,, \\
\label{position-pole-alpha_n-1-L1}
L_{n-1}&:& \ -\frac{i \epsilon}{\beta_1+\cdots +\beta_{n-1}+x+\xi}\,, \\
\label{position-pole-alpha_n-1-R1}
R_{n-1}&:& \ -\frac{i \epsilon}{\beta_1+\cdots +\beta_{n-1}+x-\xi} \,.
\eqa
and leads to the
$\beta_{n-1}$ integral of the form
\beq
\label{results-beta_n-1-spectrum}
-2 \pi i \int^{\xi-x-\beta_1 - \cdots -\beta_{n-2}}_0 d \beta_{n-1} \, {\rm Res}_{\alpha_{n-1}}\,.
\eq
The last stage is achieved by considering the $\alpha_{n}$ poles. Since the additional fermionic propagator which joins the two photon vertices has a pole which position along the imaginary axis is the same as the one of $R_n$, their position is governed  by
\beqa
\label{position-pole-alpha_n-n}
k_{n}^2&:& \ -\frac{i \epsilon}{\beta_{n}}\,, \\
\label{position-pole-alpha_n-Ln}
L_{n}&:& \ -\frac{i \epsilon}{\beta_1+\cdots +\beta_{n}+x+\xi}\,, \\
\label{position-pole-alpha_n-Rn}
R_n&:& \ -\frac{i \epsilon}{\beta_1+\cdots +\beta_{n}+x-\xi} \,,
\eqa
and
leads to the
$\beta_{n}$ integral 
\beq
\label{results-beta_n-spectrum}
-2 \pi i \int^{\xi-x-\beta_1 - \cdots -\beta_{n-1}}_0 d \beta_{n} \, {\rm Res}_{\alpha_{n}}\,.
\eq


\subsection{Some useful integrals}
\label{Ap:integrals}

In this appendix we show that 
\beq
\left.\int^\infty_0 \frac{\log^p y}{y+\Delta} dy \right|_{\rm finite}= -\frac{1}{p+1} \log^{p+1} \Delta \,.
\label{result-integral}
\eq
Consider the integral 
\beq
I_p(\Delta) = \int^\Lambda_0 \frac{\log^p y}{y+\Delta} dy\,,
\label{def-Ip}
\eq
which can be rewritten as
\beq
I_p(\Delta) = \int^\frac{\Lambda}{\Delta}_0 \frac{\log^p (\Delta \, t)}{t+1} dt\,.
\label{scaling-Ip}
\eq
Its derivative reads
\beq
I'_p(\Delta) = -\frac{\Lambda}{\Delta^2} \frac{\log^p \Lambda}{\frac{\Delta}\Delta +1}+ \frac{p}{\Delta} \int^\frac{\Lambda}{\Delta}_0 \frac{\log^{p-1} (\Delta \, t)}{t+1} dt\,,
\label{der-Ip-1}
\eq
and thus, in the limit $\Lambda \to \infty\,,$
\beq
I'_p(\Delta) \sim - \frac{\log^p \Lambda}{\Delta}+ \frac{p}{\Delta} 
I_{p-1}(\Delta)
\,.
\label{dev-der-Ip}
\eq
Now, based on the definition (\ref{def-Ip}), this derivative can be also expressed as 
\beq
I'_p(\Delta) = - \int^\Lambda_0 \frac{\log^{p} y}{(y+\Delta)^2} dy\,,
\label{der-Ip-2}
\eq
from which one deduces that $I'_p(\Delta)$ is UV finite. Thus, since 
 from the definition (\ref{def-Ip}) it is clear that $I_p$ is UV divergent, the divergency of $I_{p-1}$ is exactly given by 
\beq
\left. I_{p-1}(\Delta) \right|_{\rm div.} = \frac{1}{p} \log^p \Lambda\,,
\label{Ip-div}
\eq 
with no subleading logarithmic divergent terms, in order to compensate the first term of Eq.~(\ref{dev-der-Ip}). We can thus write symbolically
\beq
I_{p}(\Delta) =   \frac{1}{p+1} \log^{p+1} \Lambda + \left. I_{p}(\Delta)\right|_{\rm finite}\,,
\label{Ip-div-finite}
\eq 
and we now determine the dominant part of $\left.I_{p}(\Delta)\right|_{\rm finite}\,,$ in the limit $\Delta \to 0$, which we denote as $\bar{I}_{p}(\Delta)\,.$
A direct calculation shows that
\beq
I_0(\Delta)=\log \Lambda - \log \Delta
\label{res-I0}
\eq
so that 
\beq
\bar{I}_0(\Delta)=- \log \Delta\,.
\label{res-I0-bar}
\eq
Thus, using Eq.~(\ref{dev-der-Ip}), this leads to
\beq
I'_1(\Delta) \sim- \frac{\log \Delta}{\Delta} \,,
\label{res-I'1}
\eq
and based on Eq.~(\ref{Ip-div-finite}) we  obtain
\beq
\bar{I}_1(\Delta) \sim  - \frac{\log^2 \Delta}2\,.
\label{res-I1-bar}
\eq
The same analysis gives
\beq
I'_2(\Delta) \sim -\frac{\log^2 \Delta}{\Delta}  \,,
\label{res-I'2}
\eq
and thus 
\beq
\bar{I}_2(\Delta) = - \frac{\log^3 \Delta}3\,.
\label{res-I2-bar}
\eq
By inspection, it is thus natural to make the ansatz
\beq
\label{ansatz-Ip-bar}
\bar{I}_p(\Delta) = - c_p \, log^{p+1} \Delta\,,
\eq
where $c_0=1\,.$
Combining Eqs.~(\ref{dev-der-Ip}) and (\ref{ansatz-Ip-bar}), we obtain the relation
\beq
(p+1) \, c_p = p \, c_{p-1}\,,
\label{eq-cp}
\eq
and thus
\beq
c_p = \frac{1}{p+1}\,,
\label{sol-cp}
\eq
which ends up the proof of Eq.~(\ref{result-integral}).


\end{appendix}


\providecommand{\href}[2]{#2}\begingroup\raggedright\endgroup

\end{document}